\documentclass[prb,aps,showpacs,floatfix,floats,nobalancelastpage,twocolumn]{revtex4-1}
\usepackage{graphicx}
\usepackage{graphics}
\usepackage{latexsym,amsmath,amssymb,bm,euscript}
\usepackage{color}
\usepackage{dcolumn}
\usepackage{bm}
\usepackage{epstopdf}
\usepackage{times}
\usepackage{multirow}
\usepackage{subfigure}

\def\ra{\rangle}
\def\la{\langle}

\def\Hc{{\rm H.c.}}

\begin{document}

\title{Distinguishing particle-hole conjugated Fractional Quantum Hall states using quantum dot mediated edge transport}
\author{Hsin-Hua Lai}
\affiliation{National High Magnetic Field Laboratory, Florida State University, Tallahassee, Florida 32310, USA}
\author{Kun Yang}
\affiliation{Physics Department and National High Magnetic Field Laboratory, Florida State University, Tallahassee, Florida 32306, USA}
\date{\today}

\begin{abstract}
We study theoretically edge transport of a fractional quantum Hall liquid, in the presence of a quantum dot inside the Hall bar with well controlled electron density and Landau level filling factor $\nu$, and show that such transport studies can help reveal the nature of the fractional quantum Hall liquid. In our first example we study the $\nu=1/3$ and $\nu=2/3$  liquids in the presence of a $\nu=1$ quantum dot.
When the quantum dot becomes large, its edge states join those of the Hall bar to reconstruct the edge states configuration. Taking randomness around the edges into account, we find that in the disorder-irrelevant phase the two-terminal conductance of the original $\nu=1/3$ system vanishes at zero temperature, while that of the $\nu=2/3$ case remains finite. This distinction is rooted in the fact that the $\nu=2/3$ state is built upon the $\nu=1$ state. In the disorder-dominated phase, the two-terminal conductance of $\nu=1/3$ system is $(1/5)\frac{e^2}{h}$ and that of $\nu=2/3$ system is $(1/2)\frac{e^2}{h}$. We further apply the same idea to the $\nu=5/2$ system which realizes either the Pfaffian or the anti-Pfaffian states. In this case we study the edge transport in the presence of a central $\nu=3$ quantum dot. If the quantum dot is large enough for its edge states joining those of the Hall bar, in the disorder-irrelevant phase the total two-terminal conductance in the Pfaffian case is $G^{Pf}_{tot}\rightarrow 2 \frac{e^2}{h}$ while that of anti-Pfaffian case is higher but not universal, $G^{aPf}_{tot}> 2 \frac{e^2}{h}$. This difference can be used to determine which one of these two states is realized at $\nu=5/2$. In the disorder-dominated phase, however, the total two-terminal conductances in these two systems are exactly the same, $G^{Pf/aPf}_{tot}=(7/3)\frac{e^2}{h}$.
\end{abstract}
\maketitle

\section{Introduction}
Two-dimensional electron systems that exhibit fractional quantum Hall effect (FQHE) \cite{Tsui_FQHE, Prange_QHE, AMChang_RMP,Stern2008204} are among the most intriguing states of matter that support fractionalized excitations, which obey fractional or even non-Abelian statistics. The edge states \cite{Wen1990, Wen1991, Wen_edge, KF_edge, Wen_IJMPB} in the FQHE (that are protected by non-trivial topological order in the bulk) also offer a highly controlled laboratory for experimental study of quantum transport in one dimension. However, the edge state transport properties are often {\em not} sufficient to distinguish between various classes of different states that may occur at a given Landau level (LL) filling factor $\nu$. One of the candidate experimental systems for realizing a non-abelian state is the FQHE state at $\nu=5/2$ \cite{Willett_1987}. Possible non-abelian states explaining the $\nu=5/2$ plateau include the Moore-Read ``Pfaffia`` (Pf) \cite{MR_Pf} and its particle-hole conjugate counterpart, the ``anti-Pfaffian`` (aPf) \cite{SSLee_aPf, Levin_aPf}.

In the absence of Landau level mixing, the Pf state and the aPf state have exactly the same energy in the bulk due to particle-hole symmetry. These two states have very closely related bulk properties, but differ fundamentally in their edge state physics. Recent experiments involving quasi-particle tunneling between opposite edges across constrictions (or point contacts) have probed quasi-particle charge \cite{Dolev2008, Radu2008} and may have revealed signatures of non-abelian statistics \cite{Willett2010,Sanghun2011}. However, they do not allow for a sharp distinction between the Pf and aPf states. Only the quasi-particle tunneling experiment in Ref.~\onlinecite{Radu2008} is sensitive to the {\it quantitative} difference between these two states which shows in certain power-law exponents \cite{footnote1}. An experimental setup which can show {\it qualitative} differences in the signals is thus highly desired for distinguishing these two particle-hole conjugated FQHE states. It was shown that Pf and aPf edges states may have qualitatively different thermal transport properties. \cite{SSLee_aPf, Levin_aPf} More recently, Seidel and one of the present authors proposed the momentum resolved tunneling experiment to this aim \cite{Seidel2009}. These experiments have not been attempted thus far. In this paper, we propose a quantum-dot mediated edge transport experiment which can potentially help distinguish these two states.

The relation between Pf and aPf is similar to that between $\nu=1/3$ and $\nu=2/3$ states, in the sense that one is the particle-hole conjugate of the other. The $1/3$ edge consists of one chiral boson mode while $2/3$ edge consists of two chiral boson modes which are counter-propagating with respect to each other. \cite{Wen1990, Wen_edge, Wen1991, MacDonald1990, Johnson1991} Pf edge consists of two modes propagating in the same direction--a chiral boson mode plus a neutral Majorana fermion mode. \cite{Wen_1/2} aPf consists of $3$ modes--one forward chiral boson mode and two {\it backward} modes similar to Pf edges. \cite{SSLee_aPf, Levin_aPf}

We first illustrate the idea of the quantum-dot mediated edge transport on the $\nu=1/3$ and $\nu=2/3$ cases and explore their properties. For a Hall bar at filling factor $\nu=1/3$ and $\nu=2/3$ respectively, we create (through applying a proper positive gate voltage) a droplet or quantum dot in the middle of the Hall bar, with increased electron density that corresponds to $\nu=1$ in the dot, as illustrated in Figs.~\ref{Stage:onethird}(a)-(b) and Figs.~\ref{Stage:twothird}(a)-(b). There are a pair of interface modes in the $1/3$ case corresponding to the edge states of the $\nu=1$ and 1/3 QH states respectively. However, there is only one interface mode in the $\nu=2/3$ case, since the $\nu=2/3$ liquid is built on a $\nu=1$ background.

The sharp differences between these two cases are revealed when the quantum dot becomes large enough (by directly controlling the gate or by lowering the temperature) so the interface states between the quantum dot and the Hall bar merge with those of the Hall bar, shown in Fig.~\ref{Stage:onethird}(e) and Fig.~\ref{Stage:twothird}(e). In this regime, we find that if the edges are in {\it disorder-irrelevant} phase \cite{KF_randomness, KF_edge} in which  the randomness around the edge can be ignored under weak-coupling Renormalization Group (RG) analysis, the two-terminal conductance of $1/3$ case is zero in zero temperature limit and that of $2/3$ case is {\it finite} but not universal (determined by the interaction strength between the edges). On the other hand, if the edges are in {\it disorder-dominated} phase in which the random disorder strength is relevant under RG flow and can not be ignored, the two-terminal conductance of $1/3$ case is $(1/5) \frac{e^2}{h}$ and that of $2/3$ case is $(1/2)\frac{e^2}{h}$.

For the $\nu=5/2$ case, since we will focus on the half-filled second LL ($\nu=1/2$) in which the Pf or the aPf is realized, we will ignore the electrons that  completely fill the first LL (whose contribution to filling factor is $\nu=2$). Focusing on the Pf and aPf states realized at $\nu=1/2$ in the second LL, we consider creating a second LL $\nu=1$ quantum dot (total filling factor $\nu=3$) in the middle of the Pf or aPf Hall bar. There are three interface modes in the Pf case corresponding to $\nu=1$ edge and the Pf edge respectively, but there are only two interface modes in the aPf case.

Similar to the $\nu=1/3~(2/3)$ case, when the quantum dot becomes large enough so that the interface modes between the quantum dot and the Pf (aPf) merge with those of the Pf (aPf) Hall bar, the edge transport properties between these two cases may be distinguished sharply, at least for certain cases. If the edges are in the disorder-irrelevant case, the two-terminal conductance of the Pf case (ignore the contribution from the completely-filled LL which is $2\frac{e^2}{h}$) is zero at zero temperature limit while that of the aPf case is {\it finite} but not universal. In the disorder-dominated phase, unfortunately, the two-terminal conductances of the Pf and aPf are exactly the same. Motivated by the presence of this caveat, in the appendix we propose a different but related experimental setup using a line junction geometry \cite{Haug1988,KF_linejunction}, shown in Fig.~\ref{Abelian:line_junction} and Fig.~\ref{5/2:line_junction}. By controlling the middle long skinny gate, we expect the two-terminal conductance across the line junction supports an additional plateau\cite{Ito2012} before reaching its maximum value (the quantized conductance $\nu \frac{e^2}{h}$) with decreasing gate voltage in the particle-hole conjugated states such as $\nu=2/3$ and aPf. However, such plateau phases are not present in the $\nu=1/3$ case and Pf case.

This paper is organized as follows, in Sec.~\ref{Sec:edge_trans_abelian} we discuss the general properties of the quantum-dot mediated edge transport in $\nu=1/3$ and $\nu=2/3$. In Sec.~\ref{Abelian:Small droplet regime} we discuss the so called small droplet regime in which the middle quantum dot is so small that the single particle discreteness is revealed. In this regime, the systems are qualitatively the same to the point contact systems and the tunneling matrix between the top edge and the bottom edge can be obtained by a second order perturbation analysis. In Sec.~\ref{Abelian:Large droplet regime} we discuss the so called large droplet regime in which the dot is large enough so that all the tunneling events are uncorrelated and we can simply focus on each tunneling event individually. In Sec.~\ref{Abelian:Edge_recons} we discuss the so called edge reconstruction regime in which the droplet is large enough for the interface modes between the middle QH droplet and the $\nu=1/3~(2/3)$ Hall bar to merge with those of the Hall bar. We later apply the same idea for the Pf and aPf systems in Sec.~\ref{Sec:edge_trans_5/2}. We then conclude with some discussions in Sec.~\ref{Sec:summary}. In Appendex~\ref{Appendix:line_junction} we propose another experiment for distinguishing the particle-hole conjugated FQHE states using a line junction geometry \cite{KF_linejunction, Haug1988}.
\section{Quantum Dot mediated edge transport in $\nu=1/3$ and $\nu=2/3$}\label{Sec:edge_trans_abelian}
\begin{figure}[t]
\includegraphics[width=\columnwidth]{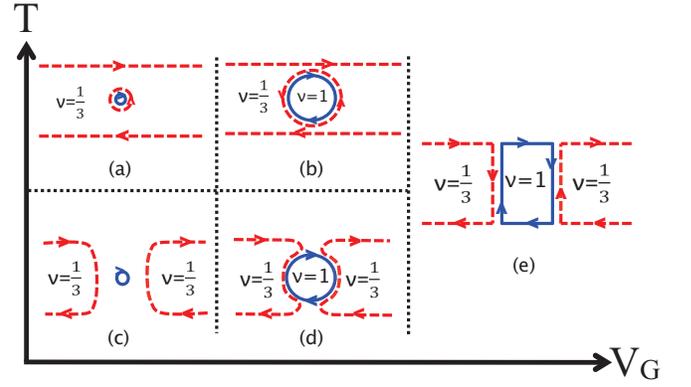}
\caption{(Color online) Schematic plots of different stages of the quantum-dot mediated edge transport system at $\nu=1/3$. The dashed red lines represent the $\nu=1/3$ edge modes and the solid blue lines represent the $\nu=1$ edge modes. The arrows represent their directions. The y-axis is the temperature and the x-axis is the voltage applied to control the central $\nu=1$ QH droplet size. We note that readers should not confuse the $V_G$ with the voltage difference between the top edge and the bottom edge of the Hall bar, $V$.
}
\label{Stage:onethird}
\end{figure}

\begin{figure}[t]
\includegraphics[width=\columnwidth]{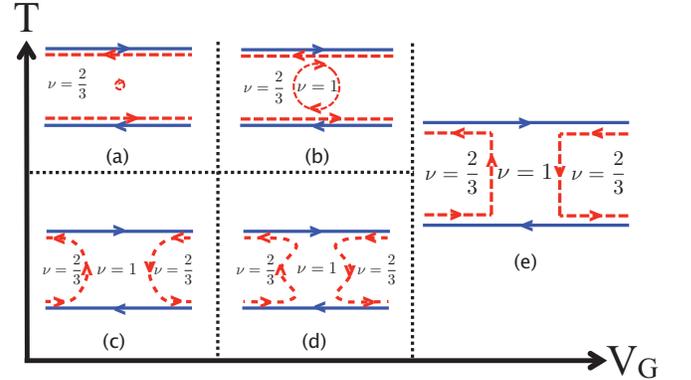}
\caption{(Color online) Schematic plots of different stages of the quantum-dot mediated edge transport system at $\nu=2/3$. The dashed red lines represent the $\nu=1/3$ edge hole modes and the solid blue lines represent the $\nu=1$ edge modes. The arrows represent their directions.
}
\label{Stage:twothird}
\end{figure}
In this section, we focus on the quantum dot mediated edge transport in the particle-hole conjugated abelian FQHE states, $\nu=1/3$ and $\nu=2/3$. The edge states of these two FQHE states have been theoretically studied for decades and are quite well understood. \cite{Wen1991,KF_randomness, KF_edge, AMChang_RMP} Under bosonization, the $\nu=1/3$ edge is described by a boson mode with charge $e/3$. The $\nu=2/3$ edge is described by two boson modes--a boson with charge $e$ similar to a $\nu=1$ edge and a {\it counter-propagating} boson with charge $-e/3$ similar to a $\nu=1/3$ hole edge. The Euclidean Lagrangians for the clean $\nu=1/3$ edge and $\nu=2/3$ edge are
\begin{eqnarray}
&& \mathcal{L}_{\frac{1}{3}} = \frac{3}{4\pi} \partial_x \phi_2(i \partial_\tau + v_2\partial_x ) \phi_2,\label{L:1/3}\\
&& \mathcal{L}_{\frac{2}{3}} = \frac{1}{4\pi} \partial_x \phi_1(i\partial_\tau+v_1\partial_x)\phi_1 + \overline{\mathcal{L}}_{\frac{1}{3}} + 2v_{12} \partial_x \phi_1 \partial_x \phi_2,\label{L:2/3}
\end{eqnarray}
where $\overline{\mathcal{L}}$ represents the Lagrangian with $\partial_\tau \rightarrow -\partial_\tau$. Under bosonization, the coarse-grained density $n_a(x) \simeq \partial_x \phi_a(x)/(2\pi)$ with $a=1,~2$. For the Lagrangian of $\nu=1/3$ edge, Eq.~(\ref{L:1/3}), the operator $e^{i\phi_2(x)}$ adds an $e/3$ quasi-particle at $x$. For the Lagrangian of $\nu=2/3$ edge, Eq.~(\ref{L:2/3}), the operator $e^{i \phi_1(x)}$ adds an $Q=e$ quasi-particle at $x$ on the $\nu=1$ edge and the operator $e^{-i \phi_2(x)}$ adds a $-e/3$ quasi-particle at $x$ on the $\nu=1/3$ edge of hole.

The schematic plots of different stages as functions of the gate voltage of the quantum dot $V_G$ and the temperature $T$ are shown in Figs.~\ref{Stage:onethird}-\ref{Stage:twothird}. When we fix the temperature and gradually increase the gate voltage, the size of the $\nu=1$ quantum dot gradually increases, which corresponds to the evolution of the configurations from left to right [(a)/(c)$\rightarrow$ (b)/(d)]. Finally the edge states of the central quantum dot and the $\nu=1/3~(2/3)$ Hall bar fuse into each other and enter the edge reconstruction regime.\cite{note2}

On the other hand, if we gradually lower the temperature while fixing the gate voltage, under RG flow, the relevant $1/3$-charged quasi-particle tunnelings increase their (effective) strength; this has an effect similar to increasing dot size, and will deplete the $\nu=1/3~(2/3)$ Hall bar into two halves, which corresponds to the evolutions of the configurations from top to bottom.  In the zero-temperature limit, we conclude that the two-terminal conductance for $\nu=1/3$ system can be either zero or $(1/5)\frac{e^2}{h}$ depending on the strength of the electron random disorder tunnelings due to the presence of randomness between the edges with counter-propagating modes. However, the two-terminal conductance of $\nu=2/3$ system is always {\it finite}. For sufficiently strong electron random disorder tunnelings, all the edges with counter-propagating modes are equilibrated and the conductance is universal and can be analytically derived to be $(1/2) \frac{e^2}{h}$.

In Sec.~\ref{Abelian:Small droplet regime}-\ref{Abelian:Edge_recons} we will first focus on the different stages when temperature $T$ is fixed and the gate voltage is gradually increased, from (a)$\rightarrow$(b)$\rightarrow$(e) in Figs.~\ref{Stage:onethird}-\ref{Stage:twothird}. Even though it is challenging to analyze the configurations in the whole regimes along the (horizontal) $V_G$-axis ($T=T_0=fixed$), there are three limiting cases that we can analyze theoretically. The three different regimes correspond to what we call the small droplet regime ((a) in Figs.~\ref{Stage:onethird}-\ref{Stage:twothird}), large droplet regime ((b) in Figs.~\ref{Stage:onethird}-\ref{Stage:twothird}), and edge reconstruction regime ((e) in Figs.~\ref{Stage:onethird}-\ref{Stage:twothird}) \cite{note2}. In each regime, we will also discuss briefly how the configurations change upon lowering the temperature. Intuitively, under RG flow, we expect to see that [(a)/(b)$\rightarrow$(c)/(d)] in Figs.~\ref{Stage:onethird}-\ref{Stage:twothird}. The configurations in edge reconstruction regime should be qualitatively the same, (e) in Figs.~\ref{Stage:onethird}-\ref{Stage:twothird}.

We note that there are two important time scales in the system with a $\nu=1$ QH droplet inside the $\nu=1/3~(2/3)$ Hall bar, (a) and (b) in Figs.~\ref{Stage:onethird}-\ref{Stage:twothird}. One of them is the time for the $-e/3$ quasi-particle staying on the central QH droplet edge before tunneling out of the droplet, $\mathtt{t_b}$. Intuitively, $\mathtt{t_b}$ should be of the same order or greater than the time scale $\mathtt{t_c}\equiv L_c/v$ with $L_c$ and $v$ being the circumference of the central QH droplet edge and its edge state velocity, $\mathtt{t_b} \geq \mathtt{t_c}$. In addition, the thermal coherence time, $\mathtt{t_T}\equiv L_T/v \simeq 1/T$, describes the time for loss of phase coherence within a single mode due to thermal smearing. With these time scales in mind we can define the regimes more explicitly.

In the small droplet regime, it is possible that $\mathtt{t_b} < \mathtt{t_T}$ and the tunneling events are coherent. If the energy cost to add an $1/3$-charged quasi-particle is much higher than the tunneling matrix elements, we can perform a second-order perturbative calculation to obtain the effective direct backscattering between the top and bottom QH bar edge. If the energy cost is comparable or somehow lower, the central QH droplet can be essentially ignored by introducing a direct backscattering matrix element between top and bottom edges. In this work, we assume the energy cost is always higher and discuss the first case in Sec.~\ref{Abelian:Small droplet regime}.

The second regime is the large droplet regime in which $\mathtt{t_b} > \mathtt{t_T}$. In this regime, the tunneling events between the QH bar edges and the central QH droplet edge are incoherent, and the dot can be viewed as an independent thermodynamic system (or lead) with well-defined temperature (assumed to be the same as the rest of the system) and chemical potential. The latter is determined by the condition that tunneling currents from top edge to central edge and from central edge to bottom edge are conserved, ($I_{t\rightarrow c} = I_{c \rightarrow b}$). In principle, we can simply calculate the tunneling current between top edge and central edge and that between central edge and bottom edge separately and qualitatively these two tunneling currents have the same power-law exponents. In Sec.~\ref{Abelian:Large droplet regime} we have detailed discussions.

In the edge reconstruction regime, the edges of the central quantum dot and the $\nu=1/3~(2/3)$ Hall bar edges fuse into each other. In this regime, $1/3$-charged quasi-particles can no longer tunnel because they are not supported outside the $\nu=1/3~(2/3)$ FQHE states. Only electrons can tunnel between the edges, and dominate transport. The detail discussions are in Sec.~\ref{Abelian:Edge_recons}.
\subsection{Small droplet regime, Fig.~\ref{Stage:onethird}(a) and Fig.~\ref{Stage:twothird}(a)}\label{Abelian:Small droplet regime}
In this regime, the size of the central $\nu=1$ QH droplet is small enough so that the discreteness of energy levels is revealed. Assuming the energy level difference is of order $U_c$ which corresponds to the charging energy to add one $1/3$-charged quasi-particle in the central droplet edge, we can perform a perturbative calculation to second order in quasi-particle tunneling between the quantum dot and one of the edges, which results in an effective backscattering matrix element of the quasi-particle between the top and bottom edge of the original $\nu=1/3$ and $\nu=2/3$ Hall bar. Note that in principle there are also tunneling events involving charge $2e/3$ and $e$ particles, but since $1/3$-charged quasi-particle tunneling is the most relevant perturbation, we ignore other possible particle tunneling process and the associated charging energies. Let us first focus on the $\nu=1/3$ case and it is easy to transit to the $\nu=2/3$ case later.
\subsubsection{$\nu=1/3$ case, Fig.~\ref{Stage:onethird}(a)}
The model Hamiltonian of the whole system (top + bottom + central edges + tunneling) under bosonization is $H= \int (\mathcal{H}_t+\mathcal{H}_b)dx + H_c + H_{t-c}+H_{c-b}$ with
\begin{eqnarray}
\mathcal{H}_{t/b}= 3\pi v_{t/b} n_{t/b}^2, \label{Honethird:top_bot}
\end{eqnarray}
and
\begin{eqnarray}
&& H_{c} = U_c N_c^2,\label{Honthird:central}\\
&& H_{t-c} = t_1 \left(e^{i(\phi_c(x_c)-\phi_t(x_t))}+ \Hc \right),~\\
&& H_{c-b} = t_2 \left(  e^{i(\phi_b(x_b)-\phi_c(x_c))}+\Hc \right),\label{Honethird:tunn}~
\end{eqnarray}
where we use $\mathcal{H}$ to stand for the Hamiltonian density and $H$ to represent the Hamiltonian. The coarse-grained density under bosonization is represented as $n_{a}(x_a) \simeq \partial_x \phi_a(x_a)/(2\pi)$ with $a=t,~b,~c$. The operator $e^{i\phi_a(x_a)}$ adds an $1/3$-charged quasi-particle at $x_a \equiv 0$ on the $a=$top,~central, or bottom edge. Below, we will suppress the labeling of $x_a$. $H_{t-c}$ and $H_{c-b}$ represent the quasi-particle tunneling terms between different edges. $H_{t/b}$ is the usual chiral Luttinger Hamiltonian and $H_c$ represent the charging energy for a capacitor, with charge $(e/3)N_c$. $N_c$ is the number of $1/3$-charged quasi-particles on the central QH droplet edge and $U_c$ is the associated energy.

In this small droplet regime, the charging energy $U_c$ is much higher than $t_1, t_2$ $( t_1, t_2 \ll U_c)$. We can perform the second-order perturbation theory to obtain an effective quasi-particle tunneling Hamiltonian between the top and bottom edge,
\begin{eqnarray}
\nonumber H_{tun}^{eff} & = &  \sum_{N_c} \frac{\left| \la N_c|e^{i \phi_c}|0\ra \right|^2}{E_0 - E_{N_c}} t_1 t_2 e^{-i \phi_t}e^{i \phi_b}  + \Hc \\
\nonumber &\simeq & -\frac{\left| \la 1 | e^{i\phi_c} |0\ra\right|^2}{U_c} t_1 t_2 e^{-i\phi_t}e^{i\phi_b} +\Hc \\
&=& t_{eff} \left(e^{i (\phi_b - \phi_t)} + \Hc \right),
\end{eqnarray}
where we assume $U_c$ is large enough so that we can ignore the virtual states with two or more quasi-particles in the central droplet edge.

With the effective quasi-particle tunneling Hamiltonian between edges of the $\nu=1/3$ Hall bar, the situation is essentially the same to the quasi-particle backscattering in a point-contact system \cite{KF1992, Moon1993, Milliken1996, KF_edge}. For simplicity, we define the tunneling operator as  $A=e^{i(\phi_t - \phi_b)}$, and the backscattering current between the top and bottom edges $I_b$ is given by the Kubo formula, \cite{Wen1991}
\begin{eqnarray}\label{G:Kubo_formula}
I_B= Q t_{eff}^2 \int_{-\infty}^{\infty} dt' \theta(t')\left[ e^{i eVt'}\la [A(t'), A^{\dagger}(0)]\ra +\Hc \right],~~~
\end{eqnarray}
where $Q$ is the charge of the tunneling quasi-particle (here $Q=e/3$).

We can see that the calculation of the backscattered currents involves the multiplication of the local retarded Green's functions of the form $G_R(t) = -i \theta(t) \la e^{i(\phi(t) -\phi(0))}\ra$, which can be related to the imaginary time Green's function $\mathcal{G}(\tau)= -\la T_\tau [ e^{i(\phi(\tau) - \phi(0))} ]\ra$ in the frequency space with $G_R(\omega) = \mathcal{G}(i \omega_n \rightarrow \omega + i 0^{+})$.

With the backscattered process, the total current is reduced by the backscattered current $I_B= I - I_0$. The reduction of the conductance, $G\equiv I/V$, thus can be obtained from the backscattered current using Eq.~(\ref{G:Kubo_formula}),
\begin{equation}
G-\left( \frac{1}{3}\right) \frac{e^2}{h} \propto \left\{
\begin{array}{lr}
- t_{eff}^2 V^{-\frac{4}{3}} ,& eV>T \\
-t_{eff}^2 T^{-\frac{4}{3}},& eV<T
\end{array}
\right\}.
\end{equation}
with $\hbar\equiv 1$.

Let us now discuss how the configuration changes upon lowering the temperature. First, it is instructive to study the quasi-particle tunneling under perturbative RG flow equation. The lowest order RG flow equation can be obtained from the scaling dimension of the quasi-particle tunneling operator $e^{i(\phi_a - \phi_b)}$, with $a,~b$ being the top, center, or bottom. According to Eq.~(\ref{L:1/3}), the scaling dimension of $e^{i \phi_a}$ can be extracted to be $\Delta[e^{\pm i \phi_a}] = 1/6$. \cite{KF_randomness, KF_edge, KF1995} The RG flow equation for the quasi-particle backscattering strength $t_b$ between top and bottom edges is
\begin{eqnarray}
\frac{dt_b}{d\ell} = (1 - \frac{1}{3}) t_b=\frac{2}{3} t_b,
\end{eqnarray}
which is relevant as expected and $\ell$ is the logarithm of the length scale. The perturbative result of the conductance can also be obtained by integrating the RG flow equation until the cutoff is of order $T$ or $eV$, giving $t_{b}^{eff} \sim [t_b(\ell_{cut})]^{-2/3}\sim T^{-2/3}~(V^{-2/3})$ and $G\sim (t^{eff}_b)^2 \sim T^{-4/3} ( V^{-4/3})$. \cite{KF_edge}

At $T\rightarrow 0$, the relevant $1/3$-charged quasi-particle tunneling perturbation will deplete the $\nu=1/3$ Hall bar into two halves that are weakly connected via the small $\nu=1$ quantum dot bridge, Fig.~\ref{Stage:onethird}(c). In this case, the conductance is only contributed by the electrons tunneling from left Hall bar edge to the right Hall bar edge via an effective tunneling matrix element that arises from the second-order perturbation. The effective electron tunneling Hamiltonian is
\begin{eqnarray}
H_{int} = t_e \left(e^{i 3(\phi_L - \phi_R)} + \Hc \right),
\end{eqnarray}
where the electron creation (annihilation) operator is operated on $x_{R/L} =0 $ and $\phi_{L/R}$ represent the chiral boson field on the Left/Right $\nu=1/3$ QH bar edge. The RG flow equation is
\begin{eqnarray}
\frac{dt_e}{d\ell} = (1-3) t_e = -2 t_e.
\end{eqnarray}
The electron tunneling perturbation strength $t_e$ is irrelevant under RG flow, which indicates that the configuration, Fig.~\ref{Stage:onethird}(c), is at a stable fixed point. The conductance in this limit with fixed two-terminal voltage, Fig.~\ref{Stage:onethird}(c), possess a power-law behavior as
\begin{eqnarray}
G\propto T^4\rightarrow 0.
\end{eqnarray}
Note that above we only consider the behaviors of each phase in the perturbative regime. As for the temperature-crossover regime, since in this case the system is essentially the same to a point-contact system, we expect that the temperature-dependent crossover can be well-captured by the thermodynamic Bethe ansatz (TBA) analysis. \cite{Fendley_TBA_1/3, Fendley1995}

Now let us shift our focus on the $\nu=2/3$ case.
\subsubsection{$\nu=2/3$ case, Fig.~\ref{Stage:twothird}(a)}
The $\nu=2/3$ edges consist of two modes: A forward propagating boson mode with charge $e$, similar to a $\nu=1$ edge, and a backward propagating mode with charge $-e/3$, similar to a $\nu=1/3$ edge of holes. The Hamiltonian density for the $\nu=2/3$ edge is
\begin{eqnarray}
\nonumber \mathcal{H}_{\nu=\frac{2}{3}}&&=\pi\left[ v_1 n_1^2 + 3 v_2 n_2^2 + 2 v_{12} n_1 n_2 \right]\\
&&\hspace{-0.2cm}= \frac{1}{4\pi} \left[ v_1 (\partial_x \phi_1)^2 + 3 v_2 (\partial_x \phi_2)^2 + 2 v_{12} \partial_x \phi_1 \partial_x \phi_2 \right],~~~~~~
\end{eqnarray}
with $n_{1/2}\simeq \partial_x \phi_{1/2}/2\pi $ corresponds to the coarse-grained density of the charged quasi-particles carrying $Q= e/(-e/3)$.

The situation is similar to the $\nu=1/3$ system with the Hamiltonians of the top edge and the bottom edge are replaced with the $\nu=2/3$ edge Hamiltonian,
\begin{eqnarray}
&& H_{t/b}= \int  \mathcal{H}_{\nu=\frac{2}{3}} dx,\\
&& H_{c} = U N_c^2,\\
&& H_{t-c} = t'_1 \left(e^{i(\phi_{2,c}-\phi_{2,t})}+ \Hc \right),~\\
&& H_{c-b} = t'_2 \left(  e^{i(\phi_{2,b}-\phi_{2,c})}+\Hc \right).~
\end{eqnarray}
The creation operators $e^{i \phi_{2, a}}$ operated at $x_a=0$. Following similar discussions before, the effective tunneling matrix element of $1/3$-charged quasi-particle between the top edge and the bottom edge is
\begin{eqnarray}
t_{eff}^{\nu=2/3} \equiv t_{eff}^{'} \sim \frac{t'_1 t'_2}{U},
\end{eqnarray}
where $N_c$ is the number of the $1/3$-charged quasi-particles on the central droplet edge and $U$ is the associated charging energy.

In order for the two counter-propagating modes on $\nu=2/3$ edges to equilibrate, we need to take the random disorders around the edges into account. \cite{KF_randomness, KF_edge} With the randomness around the edge, the Hamiltonian of the random disorder tunneling between the chiral boson mode on $\nu=1$ edge and the counter-propagating boson mode on $\nu=1/3$ edge is expressed as \cite{KF_randomness, KF_edge}
\begin{eqnarray}\label{H:random_tunneling}
H_{int}=\int dx \left[ \xi(x) e^{i (\phi_1 + 3\phi_2)} + c.c.\right],
\end{eqnarray}
where $\xi(x)$ is complex. Following the discussion by Kane and Fisher\cite{KF_randomness, KF_edge}, we assume that $\xi(x)$ is a Gaussian random variable satisfying $\overline{\xi^*(x)\xi(x')}=W\delta(x-x')$. Since the perturbation due to the randomness is spatially random \cite{Giamarchi1988}, the leading RG flow equation for $W$ is
\begin{eqnarray}\label{RG:randomness}
\frac{dW}{d\ell} = (3 - 2 \Delta)W,
\end{eqnarray}
where $\Delta \equiv  (2-\sqrt{3} c)/\sqrt{1-c^2}$ is the scaling dimension of the random disorder tunneling, Eq.~(\ref{H:random_tunneling}), and $c = (2v_{12}/\sqrt{3})/(v_1 + v_2)$ with $|c|<1$.

According to the RG flow equation, there are two different phases for the $\nu=2/3$ edge---the disorder-irrelevant phase for $\Delta >3/2$, and the disorder-dominated phase for $\Delta<3/2$. In the disorder-irrelevant phase, the corresponding conductance is in general non-universal. In the disorder-dominated phase, the corresponding conductance is universal. Below, we consider the tunneling of the $1/3$-charged quasi-particle, $e^{-i\phi_2}$, between different edges and calculate the reduction of the two-terminal conductance due to the tunneling of the quasi-particles using Kubo formula, Eq.~(\ref{G:Kubo_formula}). We summarize the properties about each phase below.\\

{\it (1) Disorder-irrelevant phase}: The random disorder tunneling between the counter-propagating bosons of the $\nu=2/3$ edge is irrelevant. The backscattering of the quasi-particle reduces the net current and so reduces the two-terminal conductance. Without the backscattering, the two-terminal conductance is $(2\Delta/3) \frac{e^2}{h}$ with $\Delta$ defined in Eq.~(\ref{RG:randomness}). \cite{KF_randomness, KF_edge, KF1995} Once the backscattering of the quansi-particle is taken into account, the reduction of the conductance has the following power-law behavior
\begin{equation}
G_{2/3}-\left( \frac{2\Delta}{3} \right) \frac{e^2}{h} \propto \left\{
\begin{array}{lr}
-(t^{'}_{eff})^2 V^{\frac{2}{3\sqrt{1-c^2}}-2}&,eV > T\\
-(t^{'}_{eff})^2 T^{\frac{2}{3 \sqrt{1-c^2}}-2}&,eV<T
\end{array}
\right\}.
\end{equation}\\

{\it (2) Disorder-dominated phase}: The random disorder tunneling is relevant and the effective low-energy physics for this phase is described by two decoupled boson modes--the ``charge'' mode, $\phi_\rho$, and the ``neutral'' mode, $\phi_\sigma$. \cite{KF_randomness} The charge mode and neutral mode are defined as
\begin{eqnarray}
&& \phi_\rho = \sqrt{\frac{3}{2}}(\phi_1 +\phi_2),~\label{Def:2/3_charge_mode}\\
&& \phi_\sigma = \sqrt{\frac{1}{2}}(\phi_1 + 3 \phi_2).\label{Def:2/3_neutral_mode}
\end{eqnarray}
The $\nu=2/3$ edge Hamiltonian density along with the random disorder tunneling term can be expressed in terms of the charge and neutral modes as $H_{\nu=2/3}=\int ( \mathcal{H}_\rho + \mathcal{H}_\sigma + \mathcal{H}_{\rho\sigma}) dx$ with
\begin{eqnarray}
&& \mathcal{H}_\rho= \frac{v_\rho}{4\pi}  (\partial_x \phi_\rho)^2,\label{H:twothird_charge}\\
&& \mathcal{H}_\sigma=  \frac{v_\sigma}{4\pi} (\partial_x \phi_\sigma)^2 + [\xi(x) e^{i\sqrt{2}\phi_\sigma}+ c.c.] ,\label{H:twothird_neutral}\\
&& \mathcal{H}_{\rho \sigma} = \frac{v}{4\pi} \partial_x \phi_\rho \phi_\sigma,\label{H:twothird_mix}
\end{eqnarray}
and $v_\rho$, $v_\sigma$, and $v$ depend on the original velocities (which we do not display). In the disorder-dominated phase, the random disorder tunneling term in Eq.~(\ref{H:twothird_neutral}) and the charge-neutral coupling term, Eq.~(\ref{H:twothird_mix}), vanish under RG flow, so $\phi_\rho$ and $\phi_\sigma$ decouple.

We can see that the random disorder tunneling term is only related to the neutral mode as $e^{i\sqrt{2}\phi_\sigma}$. In the disorder-dominated phase, since charge and neutral mode decouple, we can directly extract the scaling dimension of the random disorder tunneling term, $\Delta[e^{i \sqrt{2}\phi_{\sigma}}]=1$, and similarly $\Delta[e^{i\phi_\rho}] = 1/2$. We can directly use Kubo formula, Eq.~(\ref{G:Kubo_formula}), to calculate the backscattered current in the basis of $\phi_\rho$ and $\phi_{\sigma}$, which allows us to extract the power-law behavior of the reduction of the conductance.

On the other hand, it is much easier to use the RG analysis to exact the power-law exponent. The $1/3$-charged quasi-particle backscattering term is roughly $e^{i (\phi_{2,t} - \phi_{2, b})}=e^{i [(-\phi_{\rho,t}+\phi_{\rho,b})/\sqrt{6} + (\phi_{\sigma,t}-\phi_{\sigma,b})/\sqrt{2}]}$ and its scaling dimension is $ \overline{\Delta} = 2/3$. The RG flow equation for the quasi-particle backscattering strength $t_B$ is
\begin{eqnarray}
\frac{dt_B}{d\ell} = (1- \overline{\Delta}) dt_B = \frac{1}{3} dt_B.
\end{eqnarray}
We cut off the integration of the RG flow at $T$ or $eV$ to get the effective $t_{B, eff}\sim [t_B(\ell_{cut})]^{-1/3} \sim T^{-1/3} ~(or~V^{-1/3})$. The reduction of the conductance is roughly square of the effective backscattering strength $t_{B, eff}^2$,
\begin{equation}
G_{2/3}-\left( \frac{2}{3} \right) \frac{e^2}{h} \propto \left\{
\begin{array}{lr}
- (t_{eff}^{'})^2 V^{-\frac{2}{3}} &,eV > T\\
- (t_{eff}^{'})^2 T^{-\frac{2}{3}}&,eV < T
\end{array}
\right\}.
\end{equation}

The effect of lowering temperature in this case is similar to $\nu=1/3$ case. Upon lowering the temperature, It is expected that the configuration evolves from Fig.~\ref{Stage:twothird}(a)$\rightarrow$Fig.~\ref{Stage:twothird}(c) due to the relevant $1/3$-charged quasi-particle tunneling under RG flow. In this regime, we can see that the $\nu=1$ edges are always connected from the source to drain (left to right) and a {\it finite} two-terminal conductance is expected. For the leading-order contribution to the conductance at $T\rightarrow 0$, it should be similar to that in the Figs.~\ref{Stage:twothird}(d)-(e), and we will have detail discussions in the edge construction regime in Sec.~\ref{Abelian:Edge_recons}.
\subsection{Large droplet regime, Fig.~\ref{Stage:onethird}(b) and Fig.~\ref{Stage:twothird}(b)}\label{Abelian:Large droplet regime}
In this section, we focus on the large central QH droplet case. Since the size of the central droplet is large, the energy level of the edge is continuous and is described by a usual chiral Luttinger Hamiltonian. In this regime, $\mathtt{t_c}>\mathtt{t_T}$, the tunneling events are incoherent, and the central dot can be viewed as an independent thermodynamic system with well-defined temperature and chemical potential.  The latter is determined with the assumption of current conservation from top to bottom--the tunneling current from top edge to central edge, $I_{t-c}$, is equal to that from the central edge to bottom edge, $I_{c-b}$.

We define the conductance between the top edge and central edge as $G_{tc}$ and that between the central edge and bottom edge as $G_{cb}$. The two-terminal conductance with the consideration of the quasi-particle tunneling process satisfies the relation
\begin{eqnarray}\label{Abelian:G_relation_largedroplet}
G - \nu \frac{e^2}{h} = - \left[ (G_{tc}^{-1} + G_{cb}^{-1})^{-1} \right]= - \frac{G_{tc} G_{cb}}{G_{tc} + G_{cb}},
\end{eqnarray}
where the conductances $G_{tc}$ and $G_{cb}$ can be calculated using the Kubo formula, Eq.~(\ref{G:Kubo_formula}). Since the calculations of $G_{tc}$ and $G_{cb}$ involve the same multiplication of the local correlators as $\la e^{\phi_{1,a}(0,\tau)}e^{-\phi_{1,a}(0,0)}\ra \la e^{\phi_{2,b}(0,\tau)}e^{-\phi_{2.b}(0,0)}\ra$, $G_{tc}$ and $G_{cb}$ have the same power-law behaviors. Hence, the reduction of the conductance in $\nu=1/3$ and $\nu=2/3$ show the same power-law in temperature or in the two-terminal voltage.

Similar to the discussion above, the randomness is crucial for the equilibration between two counter-propagating modes and can drive the edges with two counter-propagating modes from the disorder-irrelevant phase to the disorder-dominated phase. The systems in these two  phases show different power-law exponents.

The Hamiltonian of the whole system in $\nu=1/3$ Hall bar is similar to Eqs.~(\ref{Honethird:top_bot})-(\ref{Honethird:tunn}) with $H_c$ replaced by the chiral Luttinger Hamiltonian. We note that the interface modes between the central QH droplet and $\nu=1/3$ Hall bar are similar to the $\nu=2/3$ edge with two counter-propagating modes. Under bosonization, the Hamiltonian of the $\nu=1/3$ system is $ H = \int ( \mathcal{H}_t +\mathcal{H}_b + \mathcal{H}_c)dx + H_{t-c} + H_{c-b}$ with
\begin{eqnarray}
&& \mathcal{H}_{t/b} = \frac{3}{4\pi} v_{t/b} \left( \partial_x \phi^{t/b}_2 \right)^2,\\
&& \mathcal{H}_c = \frac{1}{4\pi} \bigg{[} v^c_1 \left( \partial_x \phi^c_1 \right)^2 + 3 v^c_2 \left( \partial_x \phi^c_2\right)^2 + 2 v^c_{12} \partial_x \phi^c_1 \partial_x \phi^c_2 \bigg{]},~~~~~~
\end{eqnarray}
and
\begin{eqnarray}
&& H_{t-c} = t_1 \left( e^{i(\phi^t_2 - \phi^c_2)} + c.c. \right),\\
&& H_{c-b} = t_2 \left( e^{i(\phi^c_1 - \phi^b_2)} +c.c. \right),
\end{eqnarray}
where the bosons $\phi_{1/2}$ carry $e/(-e/3)$ charge. The Hamiltonian for the $\nu=2/3$ case is almost the same but with $\mathcal{H}_c$ and $\mathcal{H}_{t/b}$ interchanged. We summarize the behavior of the reduction of the conductance in different phases below.\\

{\it (1) Disorder-irrelevant phase}: In this case the reduction of the conductances for $\nu=1/3$ and $\nu=2/3$ cases show the same power-law behaviors as follows
\begin{eqnarray}
\nonumber G_{1/3} - \left( \frac{1}{3}\right) \frac{e^2}{h} &\sim& G_{2/3} -  \left( \frac{2\Delta}{3} \right) \frac{e^2}{h} \propto \\
& \propto& \left\{
\begin{array}{lr}
- t^2 V^{\frac{1}{3}(1+\frac{1}{\sqrt{1-c^2}})-2} &, eV>T\\
- t^2 T^{\frac{1}{3}(1+\frac{1}{\sqrt{1-c^2}})-2} &, eV<T
\end{array}
\right\},~~~~~~
\end{eqnarray}
with $\Delta$ and $c$ defined in Eq.~(\ref{RG:randomness}) and $t$ is the tunnel matrix element which depends on $t_1$ and $t_2$. In principle, if the system is highly symmetric, $t_1 \simeq t_2$, it is expected that $t\simeq t_1 \simeq t_2$\\

{\it (2) Disorder-dominated phase}: In this case the reduction of the conductance shows
\begin{eqnarray}
\nonumber G_{1/3} -\left(\frac{1}{3}\right) \frac{e^2}{h} &\sim& G_{2/3} - \left(\frac{2}{3}\right) \frac{e^2}{h} \propto \\
&\propto& \left\{
\begin{array}{lr}
-t^2 V^{-1} &, eV>T\\
-t^2 T^{-1} &, eV<T
\end{array}
\right\}.~~
\end{eqnarray}

Below, we will briefly discuss how the configurations in the large central droplet regime change upon lowering the temperature while fixing the two-terminal voltage $V$. Under RG flow, it is expected that the relevant $1/3$-charged quasi-particle tunnelings deplete the $\nu=1/3$ edges into two halves. Intuitively, we expect that the configurations will evolve from (b)$\rightarrow$(d) in Figs.~\ref{Stage:onethird}-\ref{Stage:twothird} for $\nu=1/3$ and $\nu=2/3$. We will discuss each case separately below.
\subsubsection{$\nu=1/3$ in the large droplet regime at $T\rightarrow 0$, Fig.~\ref{Stage:onethird}(d)}
Taking the randomness into account, we list the general behaviors in each case below. We note that in this regime the system is qualitatively the same to that in the edge reconstruction regime, Fig.~\ref{Stage:onethird}(d), and we will algebraically derive the two-terminal conductance at zero temperature limit in Sec.~\ref{Abelian:Edge_recons}.\\

{\it (1) Disorder-irrelevant phase}: In this phase, the two-terminal conductance is contributed by the electrons tunneling either directly from the left $\nu=1/3$ Hall bar edge to the right $\nu=1/3$ Hall bar edge via the locally point-contact-like geometry on the top and bottom, or via a two-step process from the left $\nu=1/3$ Hall bar edge to the central droplet and then to the right $\nu=1/3$ Hall bar edge. Since the direct electron tunneling between the left and right $\nu=1/3$ QH droplet is {\it less} relevant than that between the left (right) $\nu=1/3$ QH bar edges and the central $\nu=1$ QH droplet, we expect that the two-terminal conductance is mainly determined by the electrons tunneling between the $\nu=1/3$ QH bar edges and $\nu=1$ QH droplet, which shows the power-law exponent, $G\propto T^2\rightarrow 0$ (We ignore possible correction due to the presence of the Coulomb interaction).\\

{\it (2) Disorder-dominated phase}: In this phase, because of the equilibration between the $\nu=1$ edge and the $\nu=1/3$ edge, the composite edges consisting of a $\nu=1$ mode and a $\nu=1/3$ mode are actually similar to the $\nu=2/3$ edges in the strong-coupling phase with decoupled charge and neutral modes as discussed before. The two-terminal conductance is universal and {\it finite}. The two-terminal conductance is the same to that in Fig.~\ref{Stage:onethird}(e) and we leave the detailed analysis in Sec.~\ref{Abelian:Edge_recons}.

\subsubsection{$\nu=2/3$ in the large droplet regime at $T\rightarrow 0$, Fig.~\ref{Stage:twothird}(d)}
In this case, we can see that the $\nu=1$ edges are not altered and the two-terminal conductance is expected to be {\it finite}. The system in this limit is qualitatively the same as that in the edge reconstruction regime, Fig.~\ref{Stage:twothird}(e). The randomness is crucial for the system to show a universal two-terminal conductance, that we will explicitly show in Sec.~\ref{Abelian:Edge_recons}.

\subsection{Edge reconstruction regime, Fig.~\ref{Stage:onethird}(e) and Fig.~\ref{Stage:twothird}(e)}\label{Abelian:Edge_recons}
In this section, we consider the regime in which the central $\nu=1$ QH droplet is large enough so that the edge states of the $\nu=1/3~(2/3)$ Hall bars and the edge state of the central droplet fuse into each other to reconstruct the edge state configuration, Fig.~\ref{Stage:onethird}(e) and Fig.~\ref{Stage:twothird}(e).  Let us focus on the $\nu=1/3$ case first, Fig.~\ref{Stage:onethird}(e).
\subsubsection{Edge reconstruction of the $\nu=1/3$ case, Fig.~\ref{Stage:onethird}(e)} \label{Edge_recon_1/3}
The system in this regime is qualitatively the same to that in Fig.~\ref{Stage:onethird}. In order to determine the power-law behavior of the two-terminal conductance, we again need to take the random disorder electrons tunneling into account. \cite{KF_randomness, KF_edge, KF1995}

If the random disorder tunneling is irrelevant, there is no equilibration between the $\nu=1$ edge and $\nu=1/3$ edge. The two-terminal conductance is mostly contributed by the electron tunneling between the $\nu=1/3$ and $\nu=1$ edges, which can be calculated using Kubo formula, Eq.~(\ref{G:Kubo_formula}. Besides, the Coulomb interactions between these two edges can also modify the power-law exponents.

If the random disorder tunneling is relevant, the $\nu=1$ edge and $\nu=1/3$ edge are equilibrated and can not be treated independently. The effective low-energy description of this strong-coupling phase is similar to that discussed in Ref.~\onlinecite{KF_randomness} in terms of two decoupled boson modes called charge mode and neutral mode. At this fixed point, the two-terminal conductance is expected to be $G=(1/5) \frac{e^2}{h}$, which we will derive below. We now start to discuss the properties of each case below. \\

{\it (1) Disorder-irrelevant phase}: In this case, since the central $\nu=1$ QH droplet is large enough, each electron tunneling event is incoherent and we can consider the central QH droplet as an independent thermodynamic system with well-defined temperature and chemical potential, as the case in large droplet regime. We assume the tunneling current from left Hall bar edge to the central droplet edge is $I_{LC}$ and that from the central droplet edge to the right Hall bar edge is $I_{CR}$. Under the assumption of current conservation, $I_{CR} = I_{LC}$. The two-terminal conductance in this regime satisfies the relation
\begin{eqnarray}
G = \left( G_{LC}^{-1} + G_{CR}^{-1} \right)^{-1}.
\end{eqnarray}
Similar to the discussion in the large droplet regime, since the calculations of $G_{LC}$ and $G_{CR}$ involve the same correlators, they have the same power-law behavior. By Kubo formula, the two-terminal conductance shows the general temperature dependence
\begin{eqnarray}
G \propto T^{2\left( \Delta - 1\right)},
\end{eqnarray}
where $\Delta$ is defined in Eq.~(\ref{RG:randomness}), which is $2$ if we ignore Coulomb interaction between neighboring edges.\\

\begin{figure}[t]
\includegraphics[width=\columnwidth]{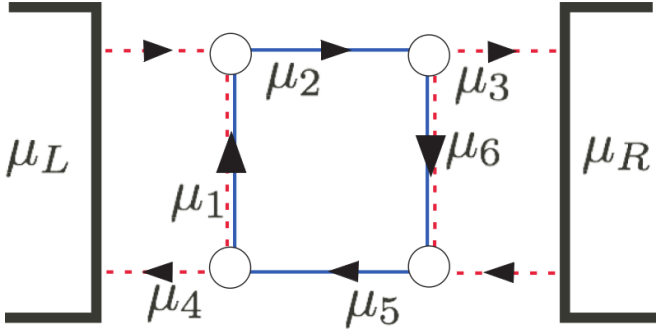}
\caption{(Color online) Closer look of the edge states network of the $\nu=1/3$ system in the edge reconstruction regime, Fig.~\ref{Stage:onethird}(e), in the disorder-dominated phase. The dashed red lines are the $\nu=1/3$ edges and the solid blue lines are the $\nu=1$ edges. The $\mu$-s represent the chemical potential of each injected currents, and the arrows represent the direction of the currents. $\mu_{R/L}$ are the source-drain voltages.
}
\label{Fig:onethird_G}
\end{figure}

{\it (2) Disorder-dominated phase}:
The edges consisting of counter-propagating modes are equilibrated, and we are able to algebraically calculate the two-terminal conductance in this case. For this end, we first show the schematic plot of the edge states network in the edge reconstruction regime, Fig.~\ref{Fig:onethird_G}. The dashed red lines represent the $\nu=1$ edges and the solid blue lines are the $\nu=1/3$ edges. Since now we are focusing on the disorder-dominated phase, a red-blue ($\nu=1,~\nu=1/3$) line junction should be treated as a $\nu=2/3$-like edge that shows the quantized $(2/3)\frac{e^2}{h}$ line conductance. The arrows represent the current directions of each line junction. $\mu_1$-$\mu_6$ represent the associated chemical potential of each current along each edge.
From current conservation, we know
\begin{eqnarray}
&& \frac{1}{3}\mu_L+\frac{2}{3}\mu_1 =\mu_2,\\
&& \mu_2 = \frac{1}{3}\mu_3 + \frac{2}{3}\mu_6, \\
&& \frac{1}{3}\mu_R + \frac{2}{3} \mu_6 = \mu_5,\\
&& \mu_5 = \frac{2}{3}\mu_1 + \frac{1}{3}\mu_4.
\end{eqnarray}
In order to solve the equations, we make the assumptions that
\begin{eqnarray}
&&\mu_2=\mu_3=\mu_6,\\
&&\mu_1=\mu_4=\mu_5.
\end{eqnarray}
These assumptions are based on the facts that currents on links 3 and 6 come from link 2 {\em only}, and currents on links 1 and 4 come from link 5 only.
With the assumptions, we obtain $\mu_1 = \mu_4 = \mu_5 = ( 3 \mu_R + 2 \mu_L)/5$ and $\mu_2 = \mu_3 =\mu_6 = (2\mu_R + 3 \mu_L)/5$. The two-terminal conductance is
\begin{eqnarray}
G=\left(\frac{\frac{1}{3}\mu_L - \frac{1}{3}\mu_4}{\mu_L - \mu_R} \right)\frac{e^2}{h} = \left(\frac{1}{5}\right) \frac{e^2}{h}.
\end{eqnarray}

In conclusion, there are two phases in the $\nu=1/3$ case. When random disorder tunneling is irrelevant, the two-terminal conductance vanishes at zero-temperature limit $G\propto T^{2(\Delta-1)}\rightarrow 0$. If the random disorder tunneling is relevant to drive each line-junction, formed by two counter-propagating $\nu=1$ and $\nu=1/3$ modes, to be disorder-dominated phase, a {\it universal} and {\it finite} two-terminal conductance $G=(1/5) (\frac{e^2}{h})$ is expected at zero temperature.
\subsubsection{Edge reconstruction of the $\nu=2/3$ case, Fig.~\ref{Stage:twothird}(e)} \label{Edge_recon_2/3}
\begin{figure}[t]
\subfigure[]{\label{Irrelevant-twothird} \includegraphics[width=\columnwidth]{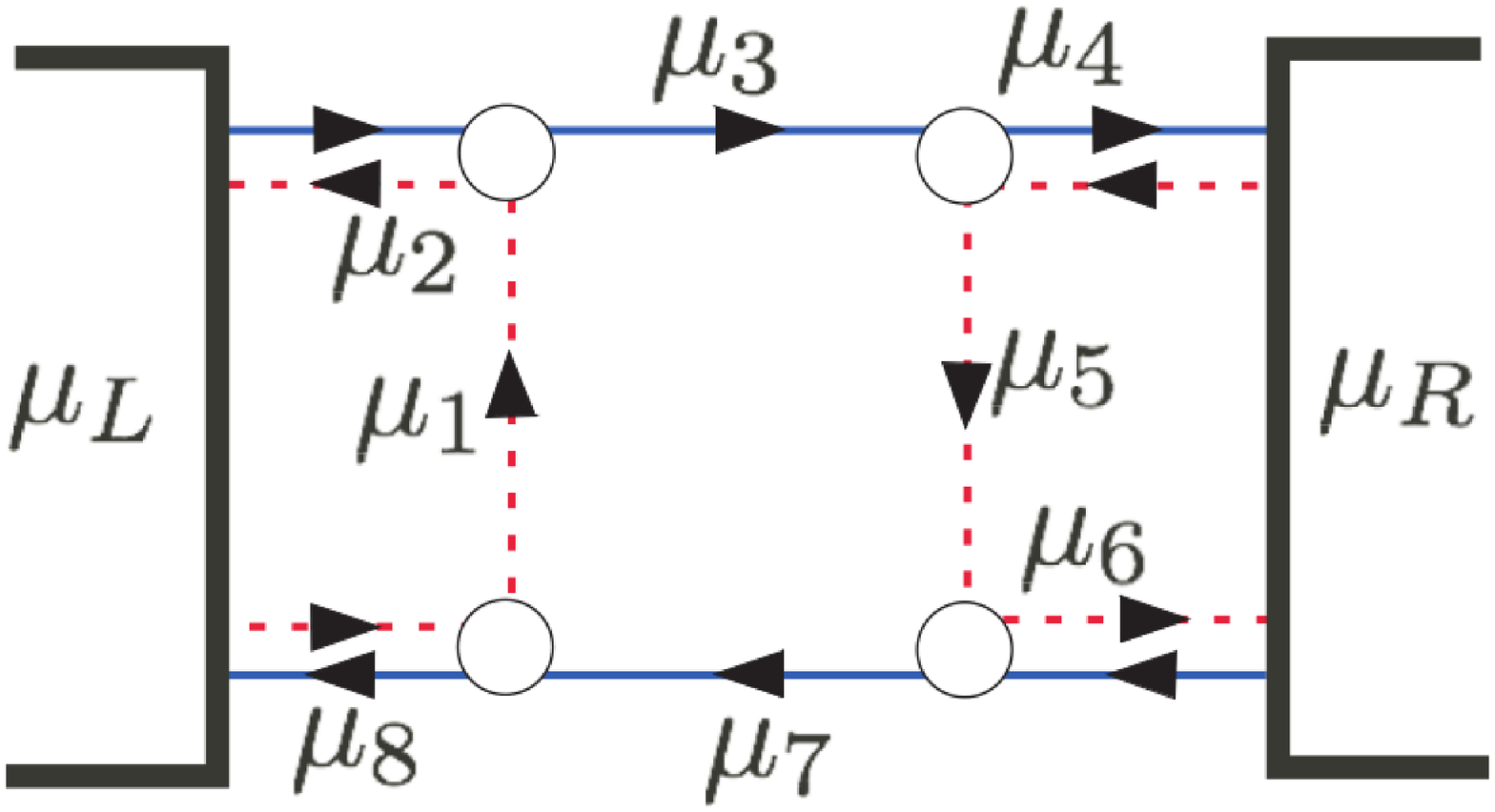}}
\subfigure[]{\label{relevant-twothird}\includegraphics[width=\columnwidth]{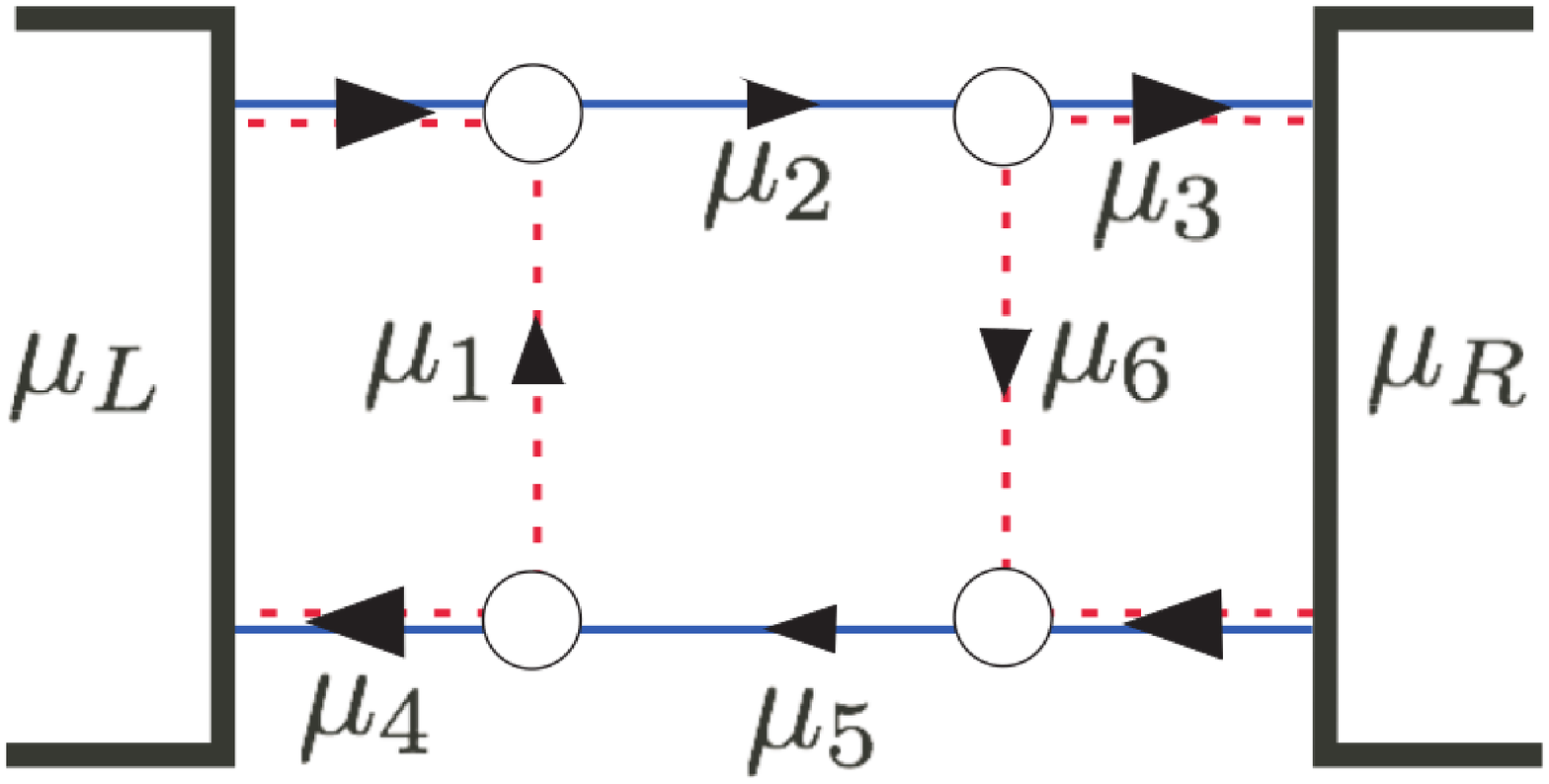}}
\caption{(Color online) Closer look of the edge states network in the $\nu=2/3$ system in the edge reconstruction regime, Fig.~\ref{Stage:twothird}(e). (a) The random disorder tunneling between each line junction formed by the $\nu=1$ mode and the counter-propagating $\nu=1/3$ mode is irrelevant. In this phase, we consider each line junction of $\nu=1$ edge and $\nu=1/3$ edge to be coupled by the Coulomb interaction but not equilibrated. In this phase, it is expected to see {\it nonuniversal} but finite conductance at zero-temperature limit. (b) The random disorder tunneling is relevant to drive the $\nu=1,~1/3$ line junctions to be in the disorder-dominated phase. In this phase, we should treat each line junction formed by $\nu=1,~1/3$ as a $\nu=2/3$-like edge in the strong-coupling phase with decoupled charge mode and neutral mode (see texts). Only the charge mode contributes to the line conductance, which is $(2/3)\frac{e^2}{h}$.}
\label{Fig:twothird_G}
\end{figure}
The system in this regime is qualitatively the same to those in Figs.~\ref{Stage:twothird}(c)-(d).
The schematic plots of the edge states network in different phases are shown in Fig.~\ref{Fig:twothird_G}. One clear feature of the $\nu=2/3$ case is that in whatever phases the system is (disorder-irrelevant or disorder-dominated), the $\nu=1$ edges are always connected from the source to drain (left to right). The two-terminal conductance in either phase should be {\it finite} at zero temperature limit. Below we will discuss each phase separately.\\

{\it (1) Disorder-irrelevant phase}:
In this case, the counter-propagating modes in $\nu=1$ edge and $\nu=1/3$ edge are not equilibrated. Though not equilibrated, these two modes are still mixed by the Coulomb interaction and the two-terminal conductance is not universal as discussed by Kane and Fisher. \cite{KF_randomness, KF_edge, KF1995} A {\it finite} but {\it nonuniversal} two-terminal conductance is expected in this case.

For an illustration of the two-terminal conductance in this phase, we ignore the Coulomb interaction between the edges and modes in $\nu=1$ and $\nu=1/3$ edges decouple. It is easy to see that the currents of the $\nu=1/3$ mode (red dashed lines) going into the system make a U-turn to go out resulting with $\mu_1=\mu_2=\mu_L$ and $\mu_5=\mu_6=\mu_R$ as far as the current conservation is concerned.

In this theoretically mode-decoupled limit, we can ignore the contribution from the $\nu=1/3$ modes and only $\nu=1$ modes (the blue line) play the role contributing to the two-terminal conductance. The conductance in this limit is
\begin{eqnarray}
G_{decoupled} = \frac{e^2}{h}.
\end{eqnarray}
Before we leave this case, we remark that if we take the Coulomb interaction into account (which is more realistic), the value of the conductance in fact depends on the strength of the coupling between different modes and is not universal, but should still be {\it finite}.\\

{\it (2) Disorder-dominated phase}: The counter-propagating modes on $\nu=1$ edge and $\nu=1/3$ edge are equilibrated to form a $\nu=2/3$-like line junction. For such a line junction, the line conductance of each line junction is universal as $(2/3) \frac{e^2}{h}$. At zero temperature, the two-terminal conductance is universal and finite. From current conservation,
\begin{eqnarray}
 && \frac{2}{3}\mu_L + \frac{1}{3}\mu_1 = \mu_2,\\
 && \mu_2 = \frac{2}{3}\mu_3 + \frac{1}{3}\mu_6,\\
 && \frac{2}{3} \mu_R + \frac{1}{3}\mu_6 = \mu_5,\\
 && \mu_5 =\frac{1}{3}\mu_1+\frac{2}{3}\mu_4.~
 \end{eqnarray}
Again we assume
\begin{eqnarray}
&& \mu_1=\mu_4=\mu_5,\\
&& \mu_2=\mu_3=\mu_6,
\end{eqnarray}
and solve the equations to get $ \mu_1 = \mu_4 =\mu_5= ( 3\mu_R + \mu_L)/4$ and $\mu_2 =\mu_3 =\mu_6 = ( \mu_R + 3\mu_L )/4$. The two-terminal current is $I = (1/2) \frac{e^2}{h} (\mu_L-\mu_R)$, which gives the two-terminal conductance
 \begin{eqnarray}
 G = \left( \frac{1}{2}\right) \frac{e^2}{h},~
 \end{eqnarray}
We note that there is also contribution from the electron tunneling between the $\nu=1/3$ edges (red dashed line) at finite temperature and it will contribute to the sub-leading term in the two-terminal conductance which is proportional to $T^{4}$ if the Coulomb interaction is ignored.
\section{Quantum Dot mediated edge transport in $\nu=5/2$: Distinguishing Pfaffian from anti-Pfaffian}\label{Sec:edge_trans_5/2}
\begin{figure}[t]
\includegraphics[width=\columnwidth]{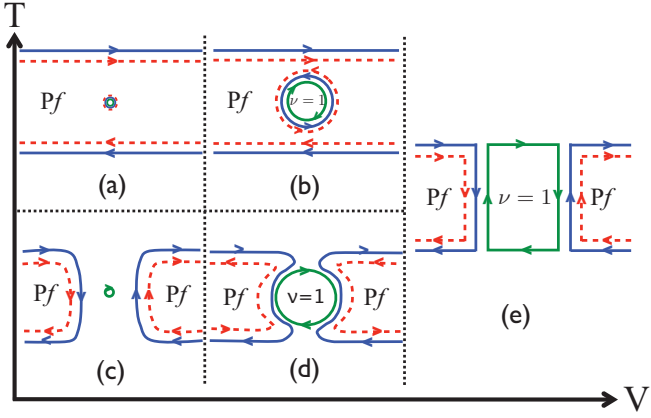}
\caption{(Color online) Schematic plots of different stages of the quantum-dot mediated edge states in the half-filled second Landau level ( $\nu=1/2$) in which the Pf state is realized. We ignore the completely filled lowest Landau level and we do not show the corresponding $\nu=2$ edges in the plots. The solid blue lines and red dashed lines represent the $\nu=1/2$ Pf edge which are described separately by chiral boson modes with charge $e/2$ and neutral Majorana fermions. The green lines represent the $\nu=1$ edge which is described by a chiral boson mode with charge $e$. Note that the edge structure of the central $\nu=1$ is similar to the aPf edge. The arrows represent the directions of each mode. The y-axis is the background temperature and the x-axis is the voltage applied to control the central $\nu=1$ (we ignore the completely-filled first LL ($\nu=2$)) QH droplet size. Again, we remind that readers not confuse the $V_G$ with the voltage difference between the top edge and the bottom edge of the Hall bar, $V$.
}
\label{Stage:Pf}
\end{figure}
\begin{figure}[t]
\includegraphics[width=\columnwidth]{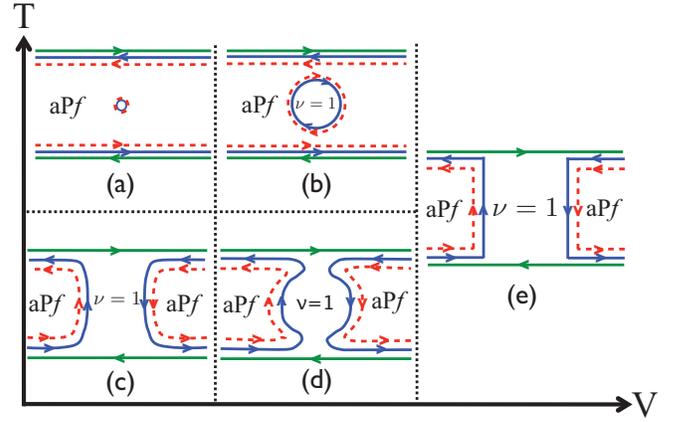}
\caption{(Color online) Schematic plots of different stages of the quantum-dot mediated edge states in the first excited half-filled Landau level ($\nu=1/2$) in which the aPf state is realized. The edge structure of the aPf edge is described by a $\nu=1$ edge mode (solid green line) and two counter-propagating modes similar to the Pf edge (a solid blue line plus a red dashed line).
}
\label{Stage:aPf}
\end{figure}
In this section, we apply the same idea to the FQHE system at filling $\nu=5/2$ in which a Pf state or aPf state is realized on the half-filled second LL ($\nu=1/2$). From now on, we will focus on the half-filled second Landau level in which the interesting Pf or aPf is realized ($\nu=1/2$) and ignore the completely-filled first Landau level ($\nu=2$) that are not significant for the following discussions. We apply a positive gate voltage in the center of the Hall bar to create a $\nu=1$ QH droplet on the second LL. The schematic plots of the $\nu=1$ quantum-dot mediated edge transport system are shown in Fig.~\ref{Stage:Pf} for Pf system and in Fig.~\ref{Stage:aPf} for aPf system.

The Euclidean Lagrangians for the Pf state and the aPf state are \cite{SSLee_aPf, Levin_aPf}
\begin{eqnarray}
&& \mathcal{L}_{Pf} = \psi(\partial_{\tau}+ i v_n \partial_x )\psi + \frac{2}{4\pi}\partial_x\phi(i\partial_\tau + v_2 \partial_x ) \phi,\label{L:Pf}\\
\nonumber && \mathcal{L}_{aPf} = \overline{\mathcal{L}}_{Pf}(\psi,\phi_2)+\frac{1}{4\pi} \partial_x \phi_1 ( i \partial_\tau + v_1 \partial_x)\phi_1 +\\
&& \hspace{1.2cm} + \frac{2}{4\pi}v_{12}\partial_x \phi_1 \partial_x \phi_2,\label{L:aPf}
\end{eqnarray}
where the $\overline{\mathcal{L}}_{Pf}$ denotes the Pf Lagrangian with $\partial_{\tau}\rightarrow -\partial_\tau$. $v_n$ is the velocity of a neutral Majorana fermion and $v_1~(v_2)$ is the velocity of a charged chiral boson in the Pf (aPf) edge. The coarse-grained density $n_a(x) \simeq \partial_x \phi_a(x)/(2\pi)$ with $a=1,~2$.

For the Pf edge, there are three types of edge quasi-particles. \cite{Milovanovic1996, Fendley2006, Bena2006, Fendley2007} An edge quasi-particle of charge $e/4$ is created by the operator $\sigma e^{i\phi/2}$ with $\sigma$ representing the Ising spin, $e/2$ quasi-particle is created by $e^{i\phi}$, and a neutral Majorana fermion $\psi$. An electron is created by the operator $\psi e^{i 2 \phi}$.

For the aPf edge, besides the edge quasi-particle creation operators (with $\phi\rightarrow \phi_2$) discussed above, an electron created on the ($\psi$, $\phi_2$) edge is represented by $\psi e^{i 2\phi_2}$ and an electron created on the $\phi_1$ edge is represented by $e^{-i\phi_1}$. \cite{SSLee_aPf, Levin_aPf} Below, we will follow the similar discussions as in the $\nu=1/3~(2/3)$ systems.
\subsection{Small droplet regime, Fig.~\ref{Stage:Pf}(a) and Fig.~\ref{Stage:aPf}(a)}\label{5/2:Small droplet regime}
In this regime, the size of the central droplet is small, so the discreteness of the energy levels is revealed. The interface edge between the central droplet and the Pf (aPf) Hall bar  is similar to the aPf (Pf) edge. In order to simplify the discussions, we model the energy costs to add an edge quasi-particle as the charging energy for a capacitor,
\begin{eqnarray}
\mathcal{H}_c = U_{\frac{1}{4}} (N^c_{1/4})^2 + U_{\frac{1}{2}} (N^c_{1/2})^2 + U_1 (N^c_e)^2,
\end{eqnarray}
where $N^c_{1/4}$, $N^c_{1/2}$, and $N^c_e$ represent the number of $e/4$ quasi-particles, of $e/2$ quasi-particles, and of electrons respectively. $U_{\frac{1}{4}}$, $U_{\frac{1}{2}}$, and $U_1$ are their charging energies associated with charging a capacitor by a certain type of the quasi-particle. In this work, we only consider the most relevant $e/4$ quasi-particle tunneling, so below we will focus on  the charging energy for adding an $1/4$-charged charged quasi-particle on the central droplet.

As discussed before, we assume the charging energy, $U_{\frac{1}{4}}$, to add an $e/4$ quasi-particle is comparably larger than the quasi-particle tunneling strength. We perform a perturbative calculation to second order to gain the effective quasi-particle backscattering matrix between the top and bottom edge of the QH bar. Similar to the case in $\nu=1/3~(2/3)$ case, after the perturbation, the system is essentially the same to the point contact system.
\subsubsection{Pfaffian case, Fig.~\ref{Stage:Pf}(a)}
The model Hamiltonian of the whole system (top and bottom Pf edges + central edge + tunneling) under bosonization is $H= \int (\mathcal{H}^{Pf}_{t} +\mathcal{H}^{Pf}_b)dx+ H_c + H_{t-c} + H_{c-b}$ with
\begin{eqnarray}
\mathcal{H}^{Pf}_{t/b}= i v_n^{t/b} \psi_{t/b} \partial_x \psi_{t/b} + 2\pi v^{t/b}(n^{t/b})^2,\label{HPf_top_bott}
\end{eqnarray}
and
\begin{eqnarray}
&& H_{c} = U_{\frac{1}{4}} (N^c_{1/4})^2,\\
&& H_{t-c} = \Gamma^{\frac{1}{4}}_1 \left( \sigma_t\sigma_c e^{i(\phi^t(x_c)-\phi^c(x_t))/2}+ \Hc \right) ,~\\
&& H_{c-b} = \Gamma^{\frac{1}{4}}_2 \left( \sigma_c \sigma_b e^{i(\phi^c(x_c)-\phi^b (x_b))/2}+\Hc \right),~
\end{eqnarray}
where $\psi$ represent the neutral Majorana fermion mode and the coarse-grained density $n^a(x_a) \simeq \partial_x \phi^a (x_a)/(2\pi)$. The operator $\sigma_a e^{i\phi_a(x_a)/2}$ creates an  $e/4$ quasi-particle at $x_a=0$ on the $a=$top, central, or bottom edge. $H_{t-c}$ and $H_{c-b}$ represent the quasi-particle tunneling terms between different edges at $x_a =0$. From now on, we will suppress the $x_a$.

The Ising spin $\sigma$ is generally nonlocal and requires additional information to be defined precisely. \cite{Milovanovic1996, Fendley2006} In our work, we assume the tunneling matrix elements $\Gamma^{\frac{1}{4}}$ are sufficiently weak so that we can ignore such complication to focus on the leading-order behavior. \cite{Bena2006} $H_c$ represents the charging energy for a capacitor with charge $(e/4)N^c_{1/4}$ and $U_{\frac{1}{4}}$ is the energy for adding a $1/4$-charged quasi-particle.

For $\Gamma^{\frac{1}{4}}_1, \Gamma^{\frac{1}{4}}_2$ $( \Gamma^{\frac{1}{4}}_1, \Gamma^{\frac{1}{4}}_2 \ll U_c)$, we perform a perturbative calculation to second order to obtain an effective quasi-particle tunneling Hamiltonian between the top and bottom QH bar edges as follows,
\begin{eqnarray}
\nonumber H_{tun}^{eff} & = &  \sum_{N^c_{1/4}} \frac{\left| \la N^c_{1/4}|\sigma_c e^{i \frac{\phi_c}{2}}|0\ra \right|^2}{E_0 - E_{N^c_{1/4}}} \Gamma^{\frac{1}{4}}_1 \Gamma^{\frac{1}{4}}_2 \sigma_b \sigma_t e^{-i \frac{\phi_t}{2}}e^{i \frac{\phi_b}{2}}  + \Hc \\
\nonumber &\simeq & -\frac{\left| \la 1 | \sigma_c e^{i\frac{\phi_c}{2}} |0\ra\right|^2}{U_{\frac{1}4}} \sigma_b \sigma_t e^{-i\frac{\phi_t}{2}}e^{i\frac{\phi_b}{2}} +\Hc \\
&=& \Gamma^{\frac{1}{4}}_{eff} \left(\sigma_b\sigma_t e^{i \frac{\phi_b - \phi_t}{2}} + \Hc \right),
\end{eqnarray}
where on the second line we assume $U_{\frac{1}{4}}$ is large enough so that we can ignore the virtual states with two or more quasi-particles in the central QH droplet edge.

This system is essentially the same to the point-contact system with weak backscattering of quasi-particles. We apply Kubo formula, Eq.~(\ref{G:Kubo_formula}), to calculate the quasi-particle backscattered current and we conclude that the reduction of the conductance due to the $e/4$ quasi-particle backscattering have the power-law behavior
\begin{equation}
G-\frac{1}{2}\frac{e^2}{h}\propto \left\{
\begin{array}{lr}
-\left( \Gamma^{\frac{1}{4}}_{eff}\right)^2 V^{-\frac{3}{2}},& eV > T \\
-\left( \Gamma^{\frac{1}{4}}_{eff} \right)^2 T^{-\frac{3}{2}},& eV<T
\end{array}
\right\}.
\end{equation}

As before, it is instructive to recast the results using RG flow analysis. For low energy physics description, we first extract the scaling dimension of the $1/4$-charged quasi-particle backscattering term. From Eq.~(\ref{L:Pf}), we know $\Delta[e^{\pm i \phi_a}] = 1/4$, and the scaling dimension of the Ising spin is $\Delta[\sigma]= 1/16$. \cite{Levin_aPf, SSLee_aPf} The total scaling dimension of the $1/4$-charged quasi-particle backscattering term is $\Delta[\sigma_a \sigma_b e^{i(\phi_a-\phi_b)/2}] = 1/4$. The RG flow equation is
\begin{eqnarray}
\frac{d\Gamma^{\frac{1}{4}}}{d\ell} = (1 - \frac{1}{4})\Gamma^{\frac{1}{4}} = \frac{3}{4}\Gamma^{\frac{1}{4}},
\end{eqnarray}
which is relevant.

Upon lowering the temperature, the relevant $1/4$-charged quasi-particle tunneling depletes the Pf Hall bar into two halves that are weakly connected via a $\nu=1$ QH droplet, Fig.~\ref{Stage:Pf}(c). Only electrons can tunnel between left and right Pf edges. Assuming the charging energy to add one electron in the central small QH droplet is much higher than the electron tunneling strength between one of the Pf Hall bar edge and the central edge, we can derive a effective electron tunneling Hamiltonian between the left Pf Hall edge and the right Pf Hall edge under perturbative calculation to second order. The effective electron tunneling term is expressed as
\begin{eqnarray}
H_{int} = \Gamma_e \left[\psi_L\psi_R e^{i2(\phi_L-\phi_R)} + \Hc \right].
\end{eqnarray}
The scaling dimension of the electron tunneling Hamiltonian is $\Delta[\psi_L \psi_R e^{i2(\phi_L-\phi_R)}] = 3$ and the RG flow equation is
\begin{eqnarray}
\frac{d\Gamma_e}{d\ell} = (1-3)\Gamma_e = -2\Gamma_e,
\end{eqnarray}
which is irrelevant. Such a configuration is a stable fixed point and the tunneling electrons can contribute to the two-terminal conductance in this regime at $T \rightarrow 0$, Fig.~\ref{Stage:Pf}(c) with the power-law behavior,
\begin{equation}
G \propto T^4\rightarrow 0.
\end{equation}

Let us shift our focus on the aPf case in the small droplet regime.
\subsubsection{anti-Pfaffian case, Fig.~\ref{Stage:aPf}(a)}
The Hamiltonian of the whole system is similar but with $H^{Pf}$ replaced by $H^{aPf}$. The Hamiltonian of the full system is $H = \int (\mathcal{H}^{aPf}_t + \mathcal{H}^{aPf}_b)dx + H_c + H_{t-c} + H_{c-b}$ with
\begin{eqnarray}
\nonumber  \mathcal{H}^{aPf}_{t/b} = && \overline{\mathcal{H}}^{Pf}_{t/b}(\psi_{t/b},\phi^{t/b}_2) + \frac{1}{4\pi}v_1^{t/b} (\partial_x \phi^{t/b}_1)^2 + \\
&&  +\frac{2}{4\pi} v^{t/b}_{12}\partial_x \phi^{t/b}_1 \partial_x \phi^{t/b}_2,
\end{eqnarray}
and
\begin{eqnarray}
&&  H_c = U_2 (N^c_2)^2,\\
&&  H_{t-c} = \Gamma^{\frac{1}{4}}_1 \left( \sigma_t\sigma_c e^{i(\phi^t_2(x_c)-\phi^c_2(x_t))/2}+ \Hc \right) ,~\\
&&  H_{c-b} = \Gamma^{\frac{1}{4}}_2 \left( \sigma_c \sigma_b e^{i(\phi^c_2(x_c)-\phi^b_2 (x_b))/2}+\Hc \right),~
\end{eqnarray}
where $\overline{\mathcal{H}}^{Pf}$ represent the Pf Hamiltonian density with the boson modes, $\phi_2$ and $\psi$, flowing backward (mathematically $\partial_\tau \rightarrow -\partial_\tau$) compared with $\phi_1$ mode.

For the aPf edge, we need to consider the tunneling between counter-propagating chiral boson modes for equilibration. Similar to the previous discussions in the $\nu=1/3~(2/3)$ case, the random disorder tunneling between the Pf channel and $\nu=1$ channel in the aPf edge is
\begin{eqnarray}
\mathcal{H}_{int} = \int dx \left[\xi'(x) \psi e^{i(\phi_1 + 2\phi_2)} \right] + c.c..
\end{eqnarray}

Following Kane, Fisher, and Polchinski's discussion, \cite{KF_randomness, Levin_aPf} we assume $\xi'$ is a complex Gaussian random variable satisfying $\la \xi'(x) \xi'^{*}(x')\ra = W' \delta(x-x')$. The scaling dimension of the above tunneling operator is
\begin{eqnarray}\label{aPf:scaling_dim}
\Delta' = \frac{1}{2} + \frac{3/2-\sqrt{2}c'}{\sqrt{1-c'^2}} \equiv \frac{1+\Lambda}{2},
\end{eqnarray}
with $\Lambda=(3-2\sqrt{2}c')/(1-c'^2)$ and $c'\equiv \sqrt{2}v_{12}/(v_1+v_2)$.

The RG flow equation for this spatially random perturbation is
\begin{eqnarray}
\frac{dW'}{d \ell} = (3-2\Delta')W'.
\end{eqnarray}
Similar to $\nu=2/3$ case, there are two phases in the aPf edge--the disorder-irrelevant phase for $\Delta'< 3/2$ and the disorder-dominated phase for $\Delta'>3/2$. The former phase possesses a nonuniversal two-terminal conductance and the latter phase possesses a universal one. Below we discuss separately the behavior of the reduction of the conductance due to the $1/4$-charged quasi-particle backscattering in these two phases.\\

{\it (1) Disorder-irrelevant phase}: The random disorder tunneling is irrelevant under RG flow and the two-terminal conductance is nonuniversal due to the lack of equilibration. Without the backscattering of $1/4$-charged quasi-particles, the two-terminal conductance can be analytically derived following the steps in Ref.~\onlinecite{KF1995}, which is $(\Lambda/2) \frac{e^2}{h}$ with $\Lambda$ defined in Eq.~(\ref{aPf:scaling_dim}).

If we take the backscattering of $1/4$-charged quasi-particle into account, the reduction of conductance in the disorder-irrelevant phase has the power-law behavior
\begin{equation}
G_{aPf}-\frac{1}{2}\Lambda \frac{e^2}{h} \propto \left\{
\begin{array}{lr}
-\left( \Gamma'^{\frac{1}{4}}_{eff} \right)^2 V^{ \frac{1}{4}+\frac{1}{4\sqrt{1-c'^2}}-2},& eV>T\\
-\left( \Gamma'^{\frac{1}{4}}_{eff} \right)^2 T^{\frac{1}{4}+\frac{1}{4\sqrt{1-c'^2}}-2},& eV<T
\end{array}
\right\},
\end{equation}
with $c'$ defined above.\\

{\it (2) Disorder-dominated phase}: The random disorder tunneling is relevant under RG flow and it flows to a stable fixed-line corresponding to $\Delta=1~(\Lambda=1/2)$.  At this stable fixed line, the coupling between the two chiral boson modes in aPf vanishes. \cite{Levin_aPf} It is more convenient to rewrite the Hamiltonian density in terms of charge and neutral modes defined as
\begin{eqnarray}
&& \phi_{\rho} = \phi_1 + \phi_2,\\
&& \phi_{\sigma} = \phi_1 + 2 \phi_2.
\end{eqnarray}
The Hamiltonian density of the chiral boson modes in terms of these charge and neutral modes (we ignore the neutral Majorana mode since it will not be altered) is
\begin{eqnarray}\
\nonumber \mathcal{H}_{U(1)} = && \frac{1}{4\pi} \left( 2 v_\rho\partial_x \phi_\rho  \partial_x \phi_\rho  + v_\sigma \partial_x \phi_{\sigma} \partial_x \phi_\sigma\right) -\\
&& - \frac{2}{4\pi} v_{\rho\sigma} \partial_x \phi_{\rho} \partial_x \phi_\sigma,\label{HaPf:U(1)}
\end{eqnarray}
where $v_{\rho}$, $v_{\sigma}$, and $v_{\rho \sigma}$ are complicated combination of $v_1$, $v_2$, and $v_{12}$ that we will not display. In this strong-coupling phase, $v_{\rho \sigma}=0$, and we can extract the scaling dimension of each chiral boson mode as
\begin{eqnarray}
&& \Delta[e^{i\phi_\rho}]\big{|}_{\Lambda=\frac{1}{2}} = \frac{1}{4},\\
&& \Delta[e^{i \phi_\sigma}]\big{|}_{\Lambda=\frac{1}{2}} = \frac{1}{2}.
\end{eqnarray}
Following the discussion of the RG analysis before, we can get the power-law behavior of the reduction of the two-terminal conductance,
\begin{equation}
G_{aPf}-\frac{1}{2}\frac{e^2}{h} \propto \left\{
\begin{array}{lr}
-\left( \Gamma'^{\frac{1}{4}}_{eff} \right)^2 V^{-1},& eV>T\\
-\left( \Gamma'^{\frac{1}{4}}_{eff} \right)^2 T^{-1},& eV<T
\end{array}
\right\}.
\end{equation}

Starting from the small QH droplet in the aPf Hall bar, we expect that the configuration evolves from Figs.~\ref{Stage:aPf}(a)$\rightarrow$\ref{Stage:aPf}(c) upon lowering the temperature. The RG flow equation suggests that the relevant $1/4$-charged quasi-particle tunneling perturbation deplete the Pf channels inside the aPf edges into two halves to form the configuration shown in Fig.~\ref{Stage:aPf}. For this configuration, the system is qualitatively the same to those in the large droplet regime at $T\rightarrow 0$, Fig.~\ref{Stage:aPf}(d), and in the edge reconstruction regime, Fig.~\ref{Stage:aPf}(e). Since the $\nu=1$ channels remain connected from the source to drain (left to right), the two-terminal conductance is expected to be finite.

For the calculation of the two-terminal conductance at zero temperature, the randomness around the edges also plays an importance role. If the random disorder tunneling between two counter-propagating chiral boson edges is irrelevant, the two-terminal conductance is expected to be {\it finite} but {\it nonuniversal}. If the random disorder tunneling is relevant, the leading order two-terminal conductance is $G= (1/3)\frac{e^2}{h}$, which we will derive explicitly in Sec.~\ref{5/2:Edge_recons}.
\subsection{Large droplet regime, Fig.~\ref{Stage:Pf}(b) and Fig.~\ref{Stage:aPf}(b)}\label{Large droplet regime}
The Hamiltonian of the whole system is similar but now the edge state of the central QH droplet can be described by the usual chiral boson Hamiltonian. We note that the interface edge between the central QH droplet and Pf Hall bar is similar to that of the aPf edge with one chiral boson mode similar to $\nu=1$ edge flowing in the opposite direction to two modes composed of one chiral boson and one neutral Majorana fermion similar to a Pf edge. On the other hand, the interface edge between the central QH droplet and aPf Hall bar is similar to that of the Pf edge.

The Hamiltonian for the whole system in the Pf case is the same as before but
with $\mathcal{H}_c$ replaced by the aPf-like Hamiltonian. The Hamiltonian of the full system is $H =  \int ( \mathcal{H}^{Pf}_t + \mathcal{H}^{Pf}_b  + \mathcal{H}_c)dx +  H_{t-c} + H_{c-b}$ with
\begin{eqnarray}
&& \mathcal{H}_{t/b}=\mathcal{H}^{Pf}_{t/b},\\
&&  \mathcal{H}_{c} = \overline{\mathcal{H}}^{Pf}_{c} +\frac{1}{4\pi} \left[ v_1^c (\partial_x \phi_1^c)^2 + 2 v^c_{12} \partial_x \phi^c_1 \partial_x \phi_2^c\right],
\end{eqnarray}
and
\begin{eqnarray}
&&  H_{t-c} = \Gamma^{\frac{1}{4}}_1 \left( \sigma_t\sigma_c e^{i(\phi^t_2 - \phi^c_2)/2}+ \Hc \right) ,~\\
&&  H_{c-b} = \Gamma^{\frac{1}{4}}_2 \left( \sigma_c \sigma_b e^{i(\phi^c_2-\phi^b_2 )/2}+\Hc \right).~
\end{eqnarray}
The Hamiltonian for the aPf case is similar but with $\mathcal{H}_c$ and $\mathcal{H}_{t/b}$ interchanged.

Similar to the previous discussion in the large droplet regime in $\nu=1/3~(2/3)$ case, we can treat each tunneling event incoherently and the central QH droplet can be treated as an independent thermodynamic system with well-defined temperature and chemical potential. The temperature is the same as that of the QH bar and the latter is determined by the assumption that the tunneling current from top QH bar edge to the central edge is the same to that from central edge to the bottom QH bar edge, $I_{t\rightarrow c} = I_{c \rightarrow b}$. We define that the conductance between the top QH bar edge and the central QH droplet edge is $G_1$ and the conductance between the central QH droplet edge and the bottom QH bar edge is $G_2$. The two-terminal conductance in this system satisfies the similar relation as in Eq.~(\ref{Abelian:G_relation_largedroplet}),
\begin{eqnarray}
G_{Pf/aPf} - \frac{1}{2}\Lambda_a\frac{e^2}{h} = - \left( G_1^{-1} + G_2^{-1} \right)^{-1},
\end{eqnarray}
where $\Lambda_{a=1,2,3}$ correspond to different cases--$\Lambda_1 = 1$ for the Pf Hall bar system, $\Lambda_2 = \Lambda$, define in Eq.~(\ref{aPf:scaling_dim}), for the aPf QH bar system in the disorder-irrelevant phase, and $\Lambda_3 = 1$ for the aPf QH bar system in the disorder-dominated phase. We will discuss each case separately below.

For the calculations of the conductances, $G_1$ and $G_2$, since they involve the same local correlators as $\la e^{i\phi_2^t(\tau)}e^{-i \phi_2^t(0)}\ra \la e^{i\phi_2^c(\tau)}e^{-i \phi_2^c(0)}\ra$, $G_1$ and $G_2$ have the same power-law behaviors. Hence, the reduction of the two-terminal conductance of Pf system is qualitatively similar to that of aPf system. We can simply focus on either $G_1$ or $G_2$, and use Kubo formula, Eq.~(\ref{G:Kubo_formula}), to extract the power-law behavior of the reduction of the two-terminal conductance.We summarize the behaviors in each case below.\\

{\it (1) Disorder-irrelevant phase}: The reduction of the two-terminal conductance shows
\begin{eqnarray}
\nonumber  G_{Pf} - \frac{1}{2} \frac{e^2}{h} &\sim& G_{aPf} - \frac{1}{2}\Lambda \frac{e^2}{h} \propto \\
& \propto& \left\{
\begin{array}{lr}
-\left( \Gamma^{\frac{1}{4}}\right)^2 V^{\frac{3}{8}+\frac{1}{8\sqrt{1-c'^2}}-2},& eV>T\\
-\left( \Gamma^{\frac{1}{4}}\right)^2 T^{\frac{3}{8}+\frac{1}{8\sqrt{1-c'^2}}-2},& eV<T\\
\end{array}
\right\},~~~~~~
\end{eqnarray}
where $\Lambda$ and $c'$ are defined in Eq.~(\ref{aPf:scaling_dim}).\\

{\it (2) Disorder-dominated phase}: The reduction of the two-terminal conductance shows
\begin{equation}
G_{Pf/aPf} - \frac{1}{2}\frac{e^2}{h} \propto \left\{
\begin{array}{lr}
-\left(\Gamma^{\frac{1}{4}}\right)^2 V^{-\frac{5}{4}},& eV>T\\
-\left(\Gamma^{\frac{1}{4}}\right)^2 T^{-\frac{5}{4}},& eV<T\\
\end{array}
\right\}.
\end{equation}

Upon lowering the temperature, it is expected that the relevant $1/4$-charged quasi-particle tunneling depletes the Pf channels of the aPf edges into halves, and the configuration evolves to Fig.~\ref{Stage:aPf}(d). The system is similar to that in the edge reconstruction regime, Fig.~\ref{Stage:aPf}(e), and we leave the detail discussions to Sec.~\ref{5/2:Edge_recons}.
\subsection{Edge reconstruction regime, Fig.~\ref{Stage:Pf}(e) and Fig.~\ref{Stage:aPf}(e)}\label{5/2:Edge_recons}
In this regime, the central QH droplet is large enough for the edge states of the central QH droplet to fuse into the edge states of the Pf (aPf) QH bar. The schematic configurations in this regime are shown in Fig.~\ref{Stage:Pf}(e) and Fig.~\ref{Stage:aPf}(e). For the Pf system, we can see that it is qualitatively the same to that with large central QH droplet at low temperature, Fig.~\ref{Stage:Pf}(d). In addition, the aPf system in this regime is similar to those in Figs.~\ref{Stage:Pf}(c)-(d) and Figs.~\ref{Stage:aPf}(c)-(d).
\subsubsection{Edge reconstruction regime of the Pfaffian case, Fig.~\ref{Stage:Pf}(e)} \label{Pf:Edge_recons}
Below we will discuss the properties of the disorder-irrelevant phase and of the disorder-dominated phase separately.

{\it (1) Disorder-irrelevant phase}: The two-terminal conductance is expected to be zero at zero temperature. For a large central droplet, the tunneling events are incoherent and we can treat each tunneling event separately. Similar to the discussion before, we assume that the electron current from the left Pf QH bar edge to the central QH droplet edge, $I_L$, is equal to that from the central QH droplet edge to the right Pf QH bar edge, $I_R$. Furthermore, we define the conductance, $G_L$, associated with $I_L$ and the conductance, $G_R$, associated with $I_R$. The two-terminal conductance of the system satisfies the relation,
\begin{eqnarray}
G = \left( G_L^{-1} + G_R^{-1}\right)^{-1}.
\end{eqnarray}
Again, since $G_L$ and $G_R$ have the same power-law behaviors, we can simply focus on one of them and use Kubo formula, Eq.~(\ref{G:Kubo_formula}), to get the power-law exponent,
\begin{equation}
G\propto \left\{
\begin{array}{lr}
t^2 V^{2\left(\Delta' -1\right)},& eV>T\\
t^2 T^{2\left(\Delta' -1\right)},& eV<T
\end{array}
\right\},
\end{equation}
with $\Delta'$ defined in Eq.~(\ref{aPf:scaling_dim}).\\

\begin{figure}[t]
\includegraphics[width=\columnwidth]{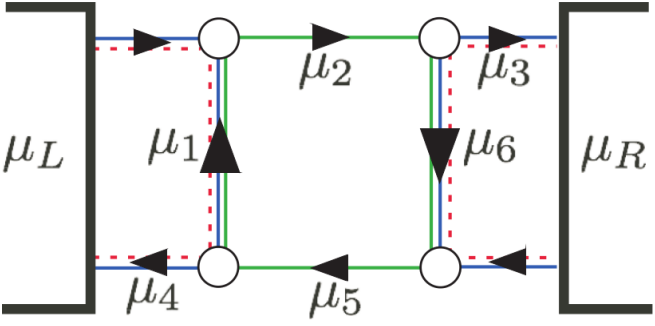}
\caption{(Color online) Closer look of the edge states network of the Pf system in the edge reconstruction regime, Fig.~\ref{Stage:Pf}(e), in the disorder-dominated phase. The counter-propagating chiral boson modes (green and blue lines) are equilibrated. The links containing such counter-propagating bosons should be treated as an equilibrated edge with one chemical potential.
}
\label{Pf:relevant}
\end{figure}

{\it (2) Disorder-dominated phase}: In this phase, the random disorder tunneling is relevant and the fixed-point theory is described by a neutral Majorana field plus two decoupled counter-propagating bosonic charge mode and bosonic neutral mode as discussed above. The edge of the two counter-propagating chiral boson modes are equilibrated to give a quantized line junction conductance $(1/2) \frac{e^2}{h}$. In order to algebraically extract the two-terminal conductance in this regime, we first focus on the schematic plot of the edge states network in this edge reconstruction regime, Fig.~\ref{Pf:relevant}.

From the current conservation, we know
\begin{eqnarray}
\frac{1}{2}\mu_L +\frac{1}{2}\mu_1 = \mu_2 \\
\mu_2 = \frac{1}{2}\mu_3 + \frac{1}{2}\mu_6 \\
\frac{1}{2}\mu_R + \frac{1}{2} \mu_6 = \mu_5 \\
\mu_5 = \frac{1}{2}\mu_1 + \frac{1}{2}\mu_4.
\end{eqnarray}
Again, we make the assumptions
\begin{eqnarray}
&& \mu_1 = \mu_4 = \mu_5,\\
&& \mu_2 = \mu_3 =\mu_6,~
\end{eqnarray}
and solve the equations to find $\mu_1=\mu_4=\mu_5=(2\mu_R+\mu_L)/3$ and $\mu_2=\mu_3=\mu_6=(\mu_R+2\mu_L)/3$. The two-terminal conductance is
\begin{eqnarray}
G_{Pf}=\frac{\frac{1}{2}\mu_L-\frac{1}{2}\mu_4}{\mu_L - \mu_R}=\frac{1}{3}\frac{e^2}{h}.
\end{eqnarray}
The total two-terminal conductance including the contribution from the completely-filled first LL ($\nu=2$) in this case is
\begin{eqnarray}
G^{Pf}_{tot} = G_{\nu=2} + G_{Pf} = \frac{7}{3}\frac{e^2}{h}.
\end{eqnarray}
\subsubsection{Edge reconstruction of the anti-Pfaffian case, Fig.~\ref{Stage:aPf}(e)} \label{aPf:Edge_recons}
\begin{figure}[t]
\subfigure[]{\label{APf:irrelevant} \includegraphics[width=\columnwidth]{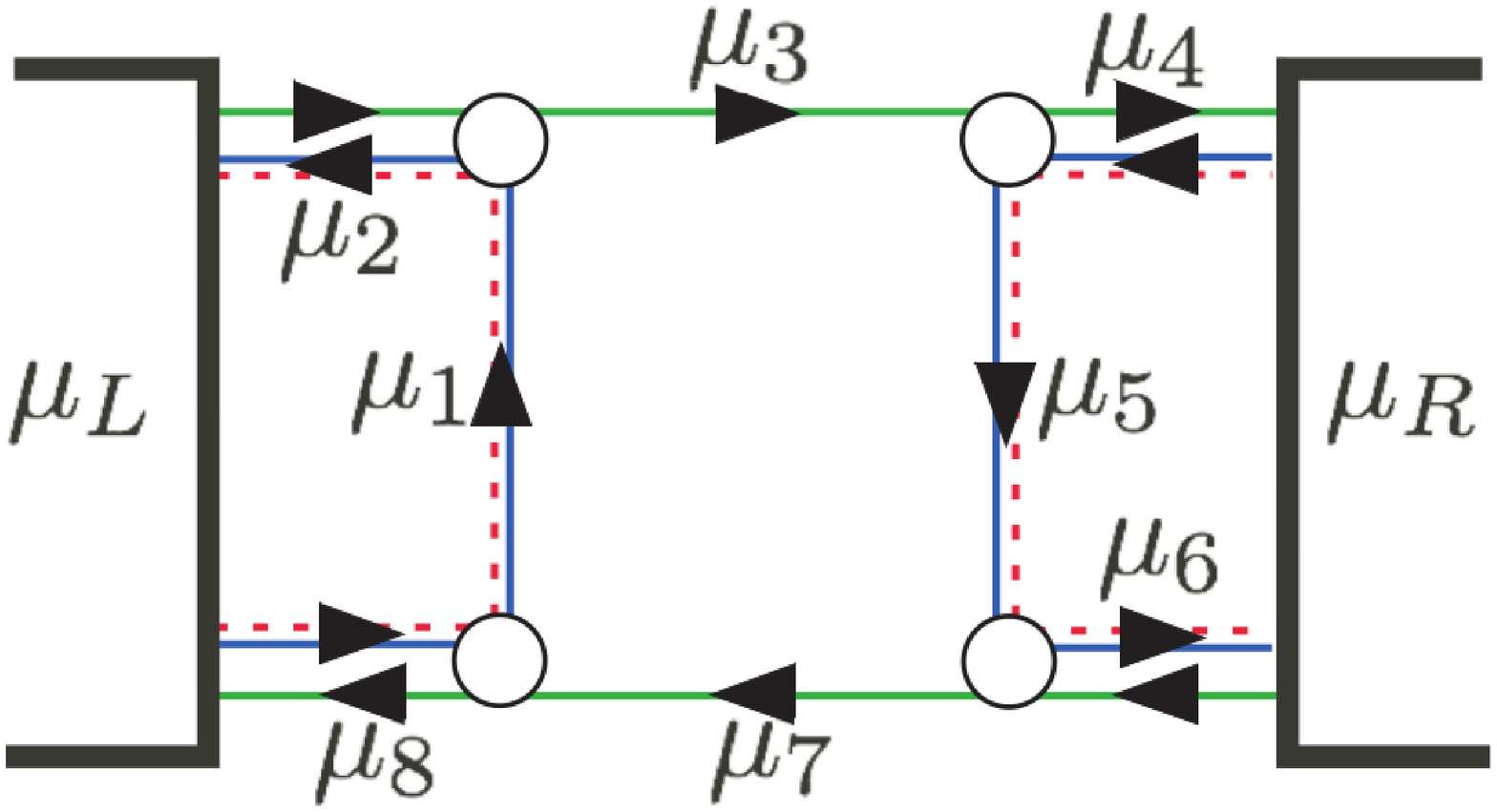}}
\subfigure[]{\label{APf:relevant}\includegraphics[width=\columnwidth]{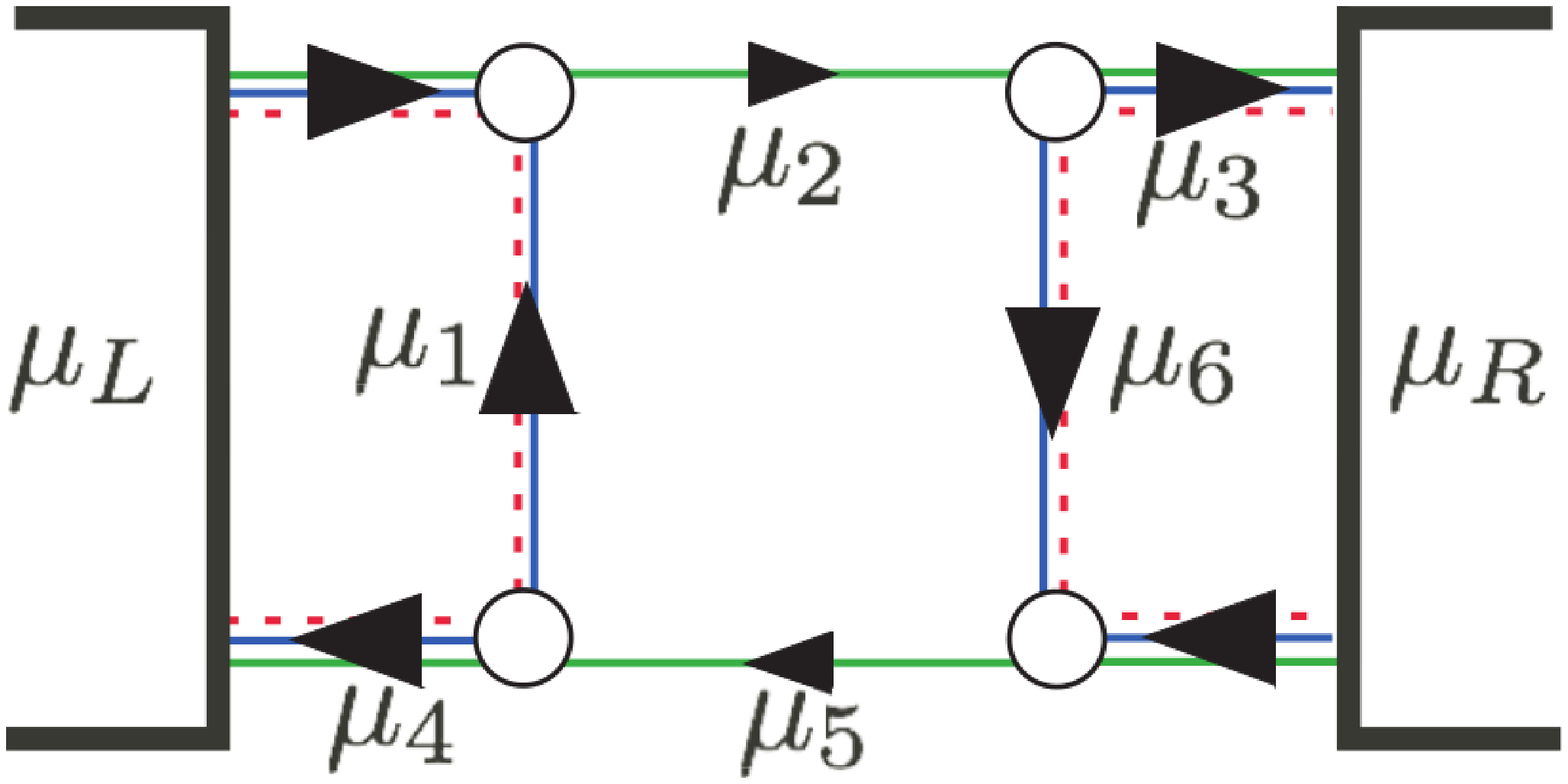}}
\caption{(Color online) The edge states network of the aPf system in the edge reconstruction regime, Fig.~\ref{Stage:aPf}(e). (a) In the disorder-irrelevant phase, the links with counter-propagating chiral boson modes (green and blue lines) are not equilibrated, and should be treated separately with different chemical potential. (b) In the disorder-dominated phase, the counter-propagating chiral boson modes are equilibrated. We should treat such links of equilibrated counter-propagating chiral boson modes as one composite edge with one chemical potential.}
\label{APf}
\end{figure}
In the aPf system, since the $\nu=1$ edges are always connected from the source to drain (left to right), the two-terminal conductance is expected to always be finite, Fig.~\ref{APf}. Taking random disorder tunneling into consideration, we will discuss the properties of the disorder-irrelevant phase and of the disorder-relevant phase separately below.\\

{\it (1) Disorder-irrelevant phase}: Due to the lack of equilibration between the two counter-propagating boson modes, the conductance is in general not universal but should be still {\it finite}. For illustration, we ignore the Coulomb interaction and each edge is decoupled from each other, the two-terminal conductance in this mode-decoupled limit is
\begin{eqnarray}
G_{decoupled} = \frac{e^2}{h}.
\end{eqnarray}\\

{\it (2) Disorder-dominated phase}: Since Majorana fermions don't contribute to the electric conductance, the situation is the same to the Pf case and we conclude
\begin{eqnarray}
G_{aPf} =G_{Pf}= \frac{1}{3}\frac{e^2}{h},
\end{eqnarray}
and the total two-terminal conductance is the same as in Pf case, $G^{aPf}_{tot} = G^{Pf}_{tot}$.

Though the electric two-terminal conductance in the disorder dominated phase here can not help distinguish the Pf state from the aPf state, it is believed that thermal conductance can be used for differing the Pf and aPf. \cite{Levin_aPf, KF_thertsp, Viola2012} However, it is difficult to measure the thermal conductance. In Appendix~\ref{Appendix:line_junction}, we propose a different but related experimental setup using a line-junction geometry, that can possibly distinguish Pf from aPf. In that system, we suggest an additional plateau signature for the particle-hole conjugated FQHE counterparts ($\nu=2/3$ and aPf) in the measurement of the two-terminal conductance across the line-junction before reaching the maximum quantized plateau at $\nu \frac{e^2}{h}$.
\section{Summary and Discussion}\label{Sec:summary}
We considered in this paper the possibility of using  quantum-dot mediated edge transports in the FQHE systems to distinguish particle-hole conjugated FQHE states such as $\nu=1/3~(2/3)$ and the Pfaffian (anti-Pfaffian) states realized at $\nu=5/2$. While unnecessary for the former (due to the difference in Hall conductance), distinguishing between Pfaffian and anti-Pfaffian states at $\nu=5/2$ is one of the pressing issues facing the quantum Hall community at present.

The difference we are trying to probe between the primary FQHE states (like 1/3 and Pfaffian) and their particle-hole conjugates (like 2/3 and anti-Pfaffian) is the latter is built on a background with integer filling. The filling factor of the quantum dot is chosen to be the the same as this background filling, as a result the dot does {\em not} suppress edge state transport even when it becomes big. On the other hand such a big dot blocks the propagation of the edge states of the primary FQHE states, and may suppress their contribution to transport.

The distinction between the primary states and their particle-hole conjugates are biggest when (random) electron tunneling across the interface between the primary state and the quantum dot is {\em irrelevant}, in which case the contribution from the primary state to two-terminal conductance vanishes in the low-temperature limit. This is a {\em stable} phase of the system. However there is another stable phase in which such electron tunneling is relevant, and the primary state does have finite contribution to the conductance. Which phase is realized will presumably depend on details of the system, and a transition between them may be observed by tuning experimental knobs like gate voltages. Such a transition is interesting in its own right.

\section{Acknowledgement}
The authors would like to thank Matthew P. A. Fisher for helpful discussions and clarification during the initial stage of this work. This research is supported by the National Science Foundation through grant No. DMR-1004545.
\appendix
\section{Distinguishing particle-hole conjugated FQHE states using line junctions}\label{Appendix:line_junction}
\begin{figure}[t]
\includegraphics[width=\columnwidth]{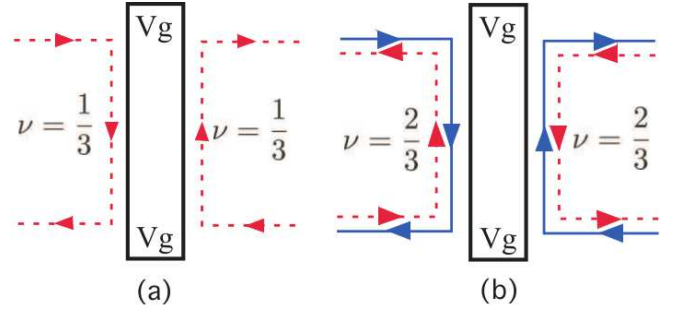}
\caption{(Color online) Schematic plots for line junctions of (a)$\nu=1/3$  and (b) $\nu=2/3$.}
\label{Abelian:line_junction}
\end{figure}

In this appendix, we suggest another experimental setup to possibly distinguish the particle-hole conjugated FQHE states using the line-junction geometry. \cite{KF_linejunction} Fig.~\ref{Abelian:line_junction} and Fig.~\ref{5/2:line_junction} show the line junctions formed by depleting $\nu=1/3~(2/3)$ and Pf (aPf) QH bars. Starting from a QH bar at certain filling $\nu$, we use a long skinny gate across the QH bar to create a line junction, with oppositely moving edge modes on either side of the gate. The strength of the electron tunneling between left QH bar and right QH bar can be varied by changing the gate voltage $V_g$. In principle, the electron tunnelings can be either relevant or irrelevant by controlling the gate voltage and there are phase transitions. \cite{KF_linejunction}

Since there are more internal edges for the particle-hole conjugated counterparts, an electron in the $\nu=2/3$ and aPf edges has more channels to tunnel across the line junction than that in the $\nu=1/3$ and Pf edges. Intuitively, there should be more phase transitions happening in $\nu=2/3$ and aPf than in the $\nu=1/3$ and Pf upon tuning the gate voltage $V_g$. The signature of the phase transitions may be extracted by measuring the two-terminal conductance across the line junction as a function of the gate voltage $V_g$.

When $V_g$ is large, the left edge is well separated from the right edge, the electron tunneling is very weak, which results in zero two-terminal conductance. Once we slowly decrease $V_g$, the electron tunneling becomes stronger. When it becomes relevant under RG flow, the relevant electron tunneling will flow to a fixed point at which the left edge and the right edge are connected. Below, we will see that for the line junctions of $\nu=2/3$ states and aPf states, there are more phase transitions than those of the $\nu=1/3$ states and Pf states. Below, we focus on the line junctions of the $\nu=1/3~(2/3)$ first and note that in this appendix, we assume that $\nu=2/3$ and aPf are in the disorder-dominated phase.
\subsection{Line junctions in $\nu=1/3$ and $\nu=2/3$: model and transition}
We will discuss the models of line junctions in $\nu=1/3$ and $\nu=2/3$ separately.
 \subsubsection{$\nu=1/3$ line junction, Fig.~\ref{Abelian:line_junction}(a)}
\begin{figure}[t]
\subfigure[]{\label{1/3:line_junction_G} \includegraphics[width=\columnwidth]{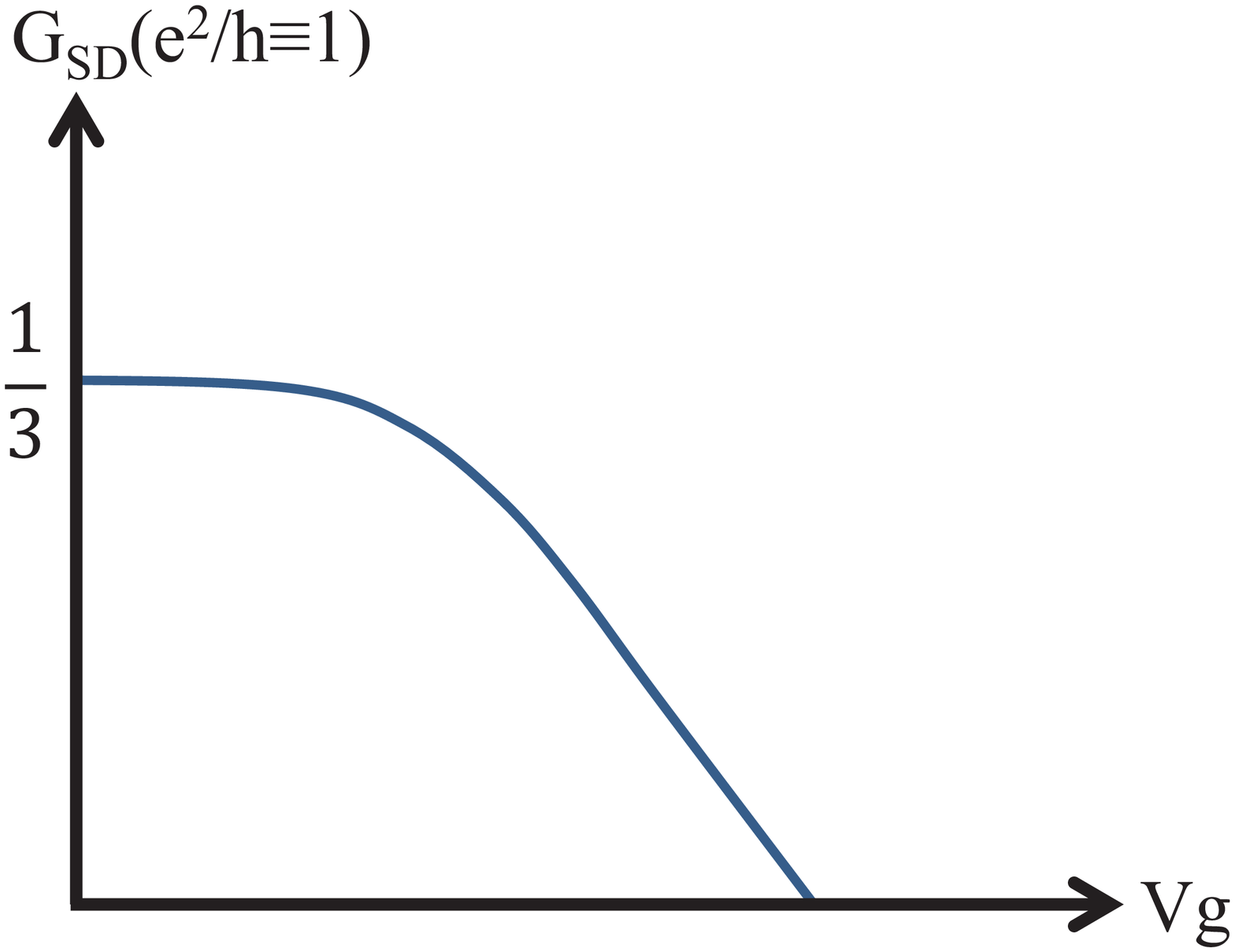}}
\subfigure[]{\label{2/3:line_junciton_G}\includegraphics[width=\columnwidth]{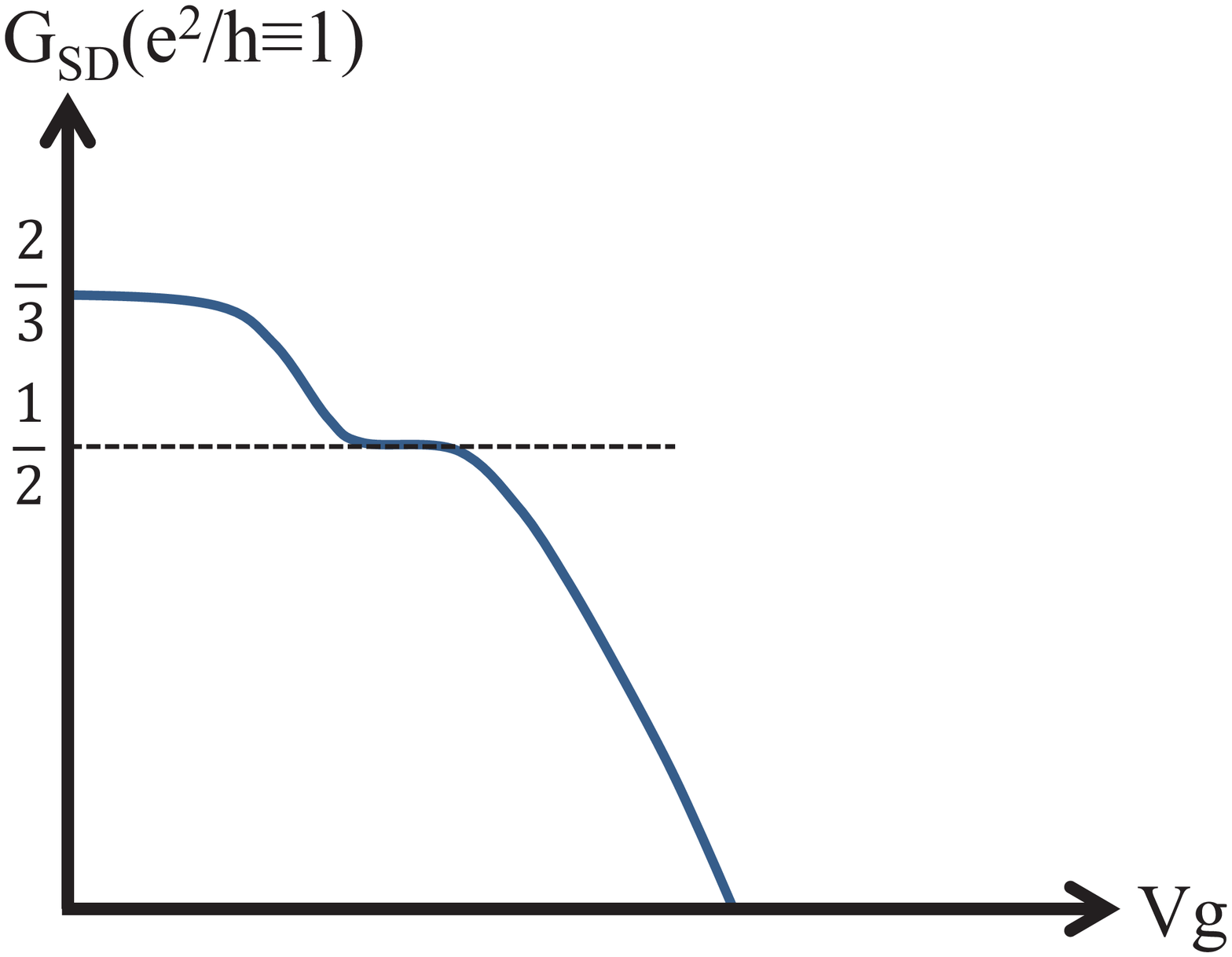}}
\caption{(Color online) Schematic plots of the conductance as a function of gate voltage $V_g$ of the long skinny gate. (a) The conductance of the $\nu=1/3$ line junction as a function of the gate voltage $V_g$. If $V_g$ is large, the left and right edges of the line junction are well separated and the conductance is zero. Upon decreasing the gate voltage, the electron tunneling becomes stronger and the conductance starts to increase. Under RG analysis, only one stable phase exists while decreasing $V_g$ which corresponds to the phase with quantized conductance $(1/3) \frac{e^2}{h}$. (b) The conductance of the $\nu=2/3$ line junction as a function of the gate voltage $V_g$. The situation is similar to (a) but upon decreasing $V_g$ there is an intermediate plateau phase with conductance, $G_{plateau}=(1/2)\frac{e^2}{h}$, smaller than the maximum quantized conductance $(2/3) \frac{e^2}{h}$.}
\label{G:abelian_line_junction}
\end{figure}
The bosonized Hamiltonian density for the $\nu=1/3$ state is given in Eq.~(\ref{Honethird:top_bot}). We will follow closely the discussions in Ref.~\onlinecite{KF_linejunction}. The Hamiltonian for a clean line junction of $\nu=1/3$ can be written in terms of right and left moving electron densities, $n_{R/L}$:
 \begin{eqnarray}
 \nonumber \mathcal{H}_0 & = & 3\pi v_0 \left(n_R^2 + n_L^2 + 2\lambda n_R n_L \right)\\
 & =& \frac{3v_0}{4\pi}\left[ (\partial_x\phi_R)^2+(\partial_x \phi_L)^2 - 2 \lambda \partial_x\phi_R \partial_x \phi_L \right],
 \end{eqnarray}
 where the last term is due to the Coulomb interaction between electrons on the left edge and on the right edge. On the second line, we re-express the density in terms of boson field $n_{R/L}=\pm \frac{1}{2\pi}\partial_x \phi_{R/L}$.

 When the gate potential is large, the two edges are well separated spatially, and the interaction strength $\lambda$ is small. As the gate potential is decreased, the modes move closer together increasing the repulsive interaction $\lambda$. The tunneling of electrons between the right and left modes under the gate becomes possible and the Hamiltonian of the electron tunneling under bosonization is
 \begin{eqnarray}
 H_1=\int dx \left( \xi(x) e^{i 3(\phi_R(x) -\phi_L(x))}+ \Hc \right),
 \end{eqnarray}
 where again $\xi(x)$ is due to the presence of random disorder and is complex. We assume that the $\xi(x)$ have a Gaussian distribution, $[\xi(x)\xi^*(x')]_{ens} = W\delta(x-x')$.

For simplicity in analysis, we define new non-chiral boson fields,
\begin{eqnarray}
\phi_{R/L} = \varphi \pm \frac{1}{3}\theta,
\end{eqnarray}
which are canonically conjugate variables,
\begin{eqnarray}
[\theta(x),\partial_x \varphi(x')]=i \delta(x-x').
\end{eqnarray}
In terms of these fields, the Hamiltonian density becomes $\mathcal{H} = \mathcal{H}_0 + \mathcal{H}_1$, with
\begin{eqnarray}
&& \mathcal{H}_0 = \frac{v}{2\pi}\left[ g(\partial_x \varphi)^2+\frac{1}{g}(\partial_x \theta)^2\right],\\
&& \mathcal{H}_1 = \xi(x)e^{i2\theta(x)} + \Hc,
\end{eqnarray}
with the renormalized velocity $v=v_0\sqrt{1- \lambda^2}$ and
\begin{eqnarray}
g=3\sqrt{\frac{1-\lambda}{1+\lambda} }.
\end{eqnarray}
The scaling dimension of the electron tunneling term is $\Delta[e^{i2\theta}]= g $. The RG flow equation to the leading order in $W$ is similar to Eq.~(\ref{RG:randomness}),
\begin{eqnarray}
\frac{dW}{d\ell} = (3 - 2 g)W.
\end{eqnarray}
When $g > 3/2$ the electron tunneling is irrelevant and the two-terminal conductance is expected to be zero in $T=0$. When $g<3/2$ the electron tunneling is relevant, the left and right $\nu=1/3$ edges fuse into each other and the two-terminal conductance is expected to be equal to the quantized Hall conductance $G=(1/3) \frac{e^2}{h}$. At $T=0$, the transition should be very sharp once crossing the critical $g_c =3/2$. However, for a realistic situation of low but finite temperature, we expect that the conductance will smoothly increase to reach the quantized conductance. The schematic plot of the conductance as a function of $V_g$ in the $\nu=1/3$ line junction is shown in Fig.~\ref{1/3:line_junction_G}.
\subsubsection{$\nu=2/3$ line junction, Fig.~\ref{Abelian:line_junction}(b)}
The Hamiltonian for the $\nu=2/3$ state in the disorder-dominated phase is given in Eqs.~(\ref{H:twothird_charge})-(\ref{H:twothird_mix}), with the coupling of the two counter-propagating modes inside the $\nu=2/3$ edge flowing to zero under RG flow. The charge mode, $\phi_\rho$, and neutral mode, $\phi_\sigma$, decouple. For this line junction, we only need to consider the Coulomb interaction between the charge modes on the opposite edges of the line junction. The Hamiltonian density for the $\nu=2/3$ line junction in disorder-dominated phase can be written in terms of boson fields as $\mathcal{H}_0 =\mathcal{H}_\rho + \mathcal{H}_\sigma$ with
\begin{eqnarray}
&& \mathcal{H}_\rho =  \frac{v_{\rho}}{4\pi}  \left[ (\partial_x \phi_{\rho L})^2+(\partial_x \phi_{\rho R})^2 - 2\lambda \partial_x \phi_{\rho L} \partial_x \phi_{\rho R} \right],~~~~\\
&&  \mathcal{H}_\sigma= \frac{v_{\sigma}}{4\pi} \left[ (\partial_x \phi_{\sigma L})^2 + (\partial_x \phi_{\sigma_R})^2\right] .~~\label{H:junction_neutral}
\end{eqnarray}
Following similar discussion above, we define new non-chiral bosons to map $\mathcal{H}_\rho$ to a Luttinger-like Hamiltonian
\begin{eqnarray}
\phi_{\rho R/L} = \varphi \pm \theta.
\end{eqnarray}
In terms of the new fields, the Hamiltonian density of the charge modes becomes
\begin{eqnarray}
\mathcal{H}_\rho = \frac{v'_\rho}{2\pi} \left[ g'(\partial_x \varphi)^2 + \frac{1}{g'}(\partial_x \theta)^2\right],\label{H:junction_rho}
\end{eqnarray}
with $v'_\rho = v_\rho \sqrt{1-\lambda^2}$ and $g'=\sqrt{\frac{1-\lambda}{1+\lambda}}$.

Starting from the well-separated line junction with large $V_g$ as shown in Fig.~\ref{Abelian:line_junction}(c), we expect the two-terminal conductance to be zero initially. Upon tuning down the gate voltage, the electrons can tunnel between the left and right $\nu=2/3$ edges. Interestingly, since there are two internal edges (channels) in each $\nu=2/3$ edge, a electron has more degrees of freedom to tunnel. An electron created in the $\nu=1$ edge is represented by the operator $e^{i\phi_{1 R/L}} = e^{i[\sqrt{\frac{3}{2}}\phi_{\rho R/L} - \frac{1}{\sqrt{2}}\phi_{\sigma R/L}]}$ and an electron created in the $\nu=1/3$ edge is represented by the operator $e^{- i 3 \phi_{2 R/L}}=e^{ i [ \sqrt{\frac{3}{2}}\phi_{\rho R/L} - \frac{3}{\sqrt{2}}\phi_{\sigma R/L}]}$ with the identity in Eqs.~(\ref{Def:2/3_charge_mode})-(\ref{Def:2/3_neutral_mode}) explicitly used. Hamiltonian of tunneling of electrons between left and right edge can be written in terms of these operators $H_1 = \int \mathcal{H}_1 dx$ with
\begin{widetext}
\begin{eqnarray}
\nonumber \mathcal{H}_1 & =& \bigg{[} \xi_1(x) e^{i (\phi_{1 L} -\phi_{1 R})} + \xi_2(x) e^{i (\phi_{1 L} + 3 \phi_{2 R})} + \xi_3 (x) e^{i 3(\phi_{2 L} - \phi_{2 R})} \bigg{]}+ \Hc \\
& = & \left[ \xi_1(x) e^{i \left( -\sqrt{6}\theta - \frac{1}{\sqrt{2}}\phi_{\sigma L} + \frac{1}{\sqrt{2}}\phi_{\sigma R}\right)} + \xi_2(x) e^{i\left( -\sqrt{6}\theta - \frac{1}{\sqrt{2}} \phi_{\sigma L} +\frac{3}{\sqrt{2}}\phi_{\sigma R}\right)} + \xi_3(x) e^{i \left( \sqrt{6}\theta +\frac{3}{\sqrt{2}}\phi_{\sigma L} -\frac{3}{\sqrt{2}} \phi_{\sigma R}\right)}\right] + \Hc,
\end{eqnarray}
\end{widetext}
where $\xi_1(x),~\xi_2(x),~\xi_3(x)$ are complex and are present due to the random disorder inside the line junction. Though there are three types of electron tunneling terms, under the assumption that all the randomnesses are in Gaussian distribution with $[\xi_a(x) \xi_a(x')]_{ens} = W_a \delta(x-x')$, with $a=1,2,3$, all of them satisfy the same RG flow equations
\begin{eqnarray}
\frac{dW_a}{d\ell} = (3 - 2 \Delta_a )W_a,
\end{eqnarray}
where $\Delta_a$ are the scaling dimensions of the electron tunneling term of type-$a$. We now examine the scaling dimension of each term to see which one is relevant and to which possible fixed point it can drive in case of relevance. The scaling dimensions of each term are
\begin{eqnarray}
&& \Delta \left[ e^{i (\phi_{1L} -\phi_{1R})} \right] = \frac{1}{2} + \frac{3}{2} g,\\
&& \Delta \left[ e^{i (\phi_{1L} + 3 \phi_{2R})} \right] = \frac{5}{2} + \frac{3}{2} g,\\
&& \Delta \left[ e^{i 3 (\phi_{2L} - \phi_{2 R})} \right] = \frac{9}{2} + \frac{3}{2} g.
\end{eqnarray}
Since $0<g <1$, the type-$1$ electron tunneling term can be relevant while type-$2$ and type-$3$ electron tunneling terms are always irrelevant. Upon decreasing the gate voltage, $\lambda$ increases while $g$ decrease. When $g>g_c = 3/2$, the electron tunneling is irrelevant, and the two-terminal conductance in this regime is expected to be zero at $T=0$. When $g< g_{c}=2/3$, the electron tunneling between $\nu=1$ edge is relevant and there is a phase transition.

\begin{figure}[t]
\includegraphics[width=\columnwidth]{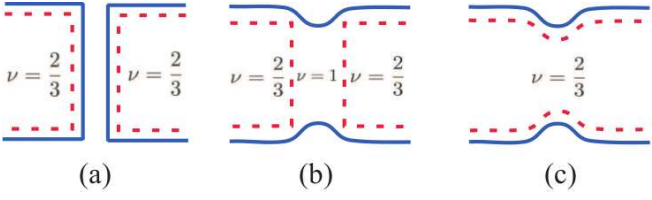}
\caption{(Color online) Schematic plots of different phases of the $\nu=2/3$ line junction. The solid blue lines are the $\nu=1$ edges and the dashed red lines are the $\nu=1/3$ edges. (a) When gate voltage $V_g$ of the central long skinny gate is large, the left and right edges are well separated with vanishing conductance. (b) When $V_g$ decreases to a threshold value, the electron tunneling between the left $\nu=1$ edge and right $\nu=1$ edge is relevant. Under RG, the $\nu=1$ edges fuse into each other and the phase is the so called plateau phase. In this phase, the configuration is the same to the edge reconstruction phase of the $\nu=2/3$ system in disorder-dominated phase, which is discussed in Sec.~\ref{Edge_recon_2/3}. The condutance in this plateau phase is expected to be $G=(1/2)\frac{e^2}{h}$. (c) When $V_g$ become too small, the $\nu=1/3$ edges also are connected to give the quantized conductance $G=(2/3) \frac{e^2}{h}$.}
\label{2/3line_junction:diffstage}
\end{figure}

For $g<g_c$, the RG flow is relevant and the physical effect is that the $\nu=1$ edges on the left and on the right fuse into each other. The configuration after the phase transition will evolve from Fig.~\ref{2/3line_junction:diffstage}(a) to Fig.~\ref{2/3line_junction:diffstage}(b).
Let us check the stability of such phase shown in Fig.~\ref{2/3line_junction:diffstage}(b). Since the width of the line junction is much narrower than its length. We can consider the electron tunneling between $\nu=1$ top edge and bottom edges to be point-contact-like and the scaling dimension of it is $\Delta[e^{i (\phi_{1 t} -\phi_{1 b})}] = 1$, which is strictly marginal.  On the other hand, the electron tunneling between the left and right $\nu=1/3$ edges is similar to the situation discussed last section about the $\nu=1/3$ line junction. If the width between the left and right $\nu=1/3$ edges is large enough, the electron tunneling is irrelevant. Therefore, the configuration in Fig.~\ref{2/3line_junction:diffstage}(b) is stable. Focusing on this phase, we recognize that the configuration here is the same to the edge reconstruction regime in $\nu=2/3$ system in the disorder-dominated phase that we discussed in Sec.~\ref{Edge_recon_2/3}. The conductance in this stage is expected to be $G_{plateau} = (1/2)\frac{e^2}{h}$.

If we further decrease the gate voltage $V_g$ to decrease the distance between left and right edges, it is expected the electron tunneling between the $\nu=1/3$ edges can become relevant to drive another phase transition which fuses the left $\nu=1/3$ and right $\nu=1/3$ edges. The configuration in this phase is shown in Fig.~\ref{2/3line_junction:diffstage}(c). In this regime, we expect the two-terminal conductance reaches its maximum which is the quantized conductance $G_{2/3} = (2/3) \frac{e^2}{h}$.

In conclusion, in the $\nu=2/3$ line junction system, if we start from a large gate voltage, $V_g$, the two-terminal conductance across the line junction is zero. Upon decreasing $V_g$, the conductance increases until hitting a plateau, with $G_{plateau} = 1/2 \frac{e^2}{h}$, that corresponds to the intermediate stable phase, Fig.~\ref{2/3line_junction:diffstage}(b). If we keep lowering $V_g$, the conductance will leave the plateau and starts to increase until reaching the quantized conductance $G_{2/3} =(2/3)\frac{e^2}{h}$, which corresponds to the phase shown in Fig.~\ref{2/3line_junction:diffstage}(c) . The two-terminal conductance as a function of $V_g$ is shown in Fig.~\ref{2/3:line_junciton_G}. The existence of the plateau phase is a crucial feature for the $\nu=2/3$.

Now, we can apply the same idea to the Pf and aPf line-junction systems.
\subsection{Line junctions in Pfaffian and anti-Pfaffian: model and transition}
In this section we will discuss the properties of the line junctions formed by either Pf states or aPf states. Much discussions in the previous section can apply here.
\begin{figure}[t]
\includegraphics[width=\columnwidth]{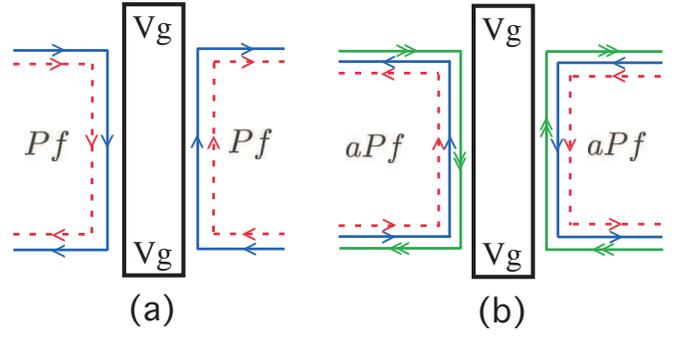}
\caption{(Color online) Schematic plots for line junctions of (a) Pf and of (b) aPf}
\label{5/2:line_junction}
\end{figure}
\subsubsection{Pfaffian line junction, Fig.~\ref{5/2:line_junction}(a)}
\begin{figure}[t]
\subfigure[]{\label{Pf:line_junction_G} \includegraphics[width=\columnwidth]{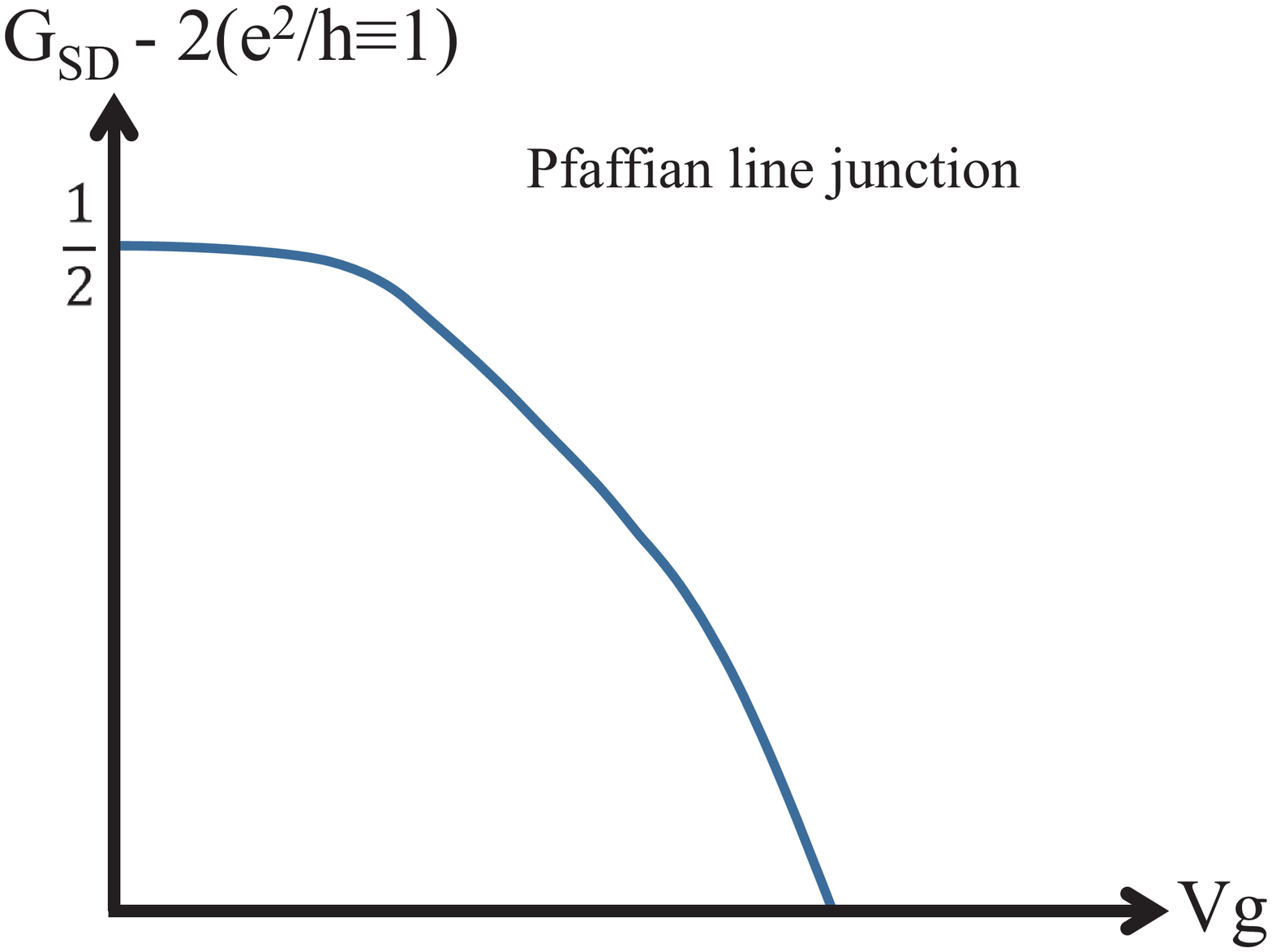}}
\subfigure[]{\label{aPf:line_junction_G}\includegraphics[width=\columnwidth]{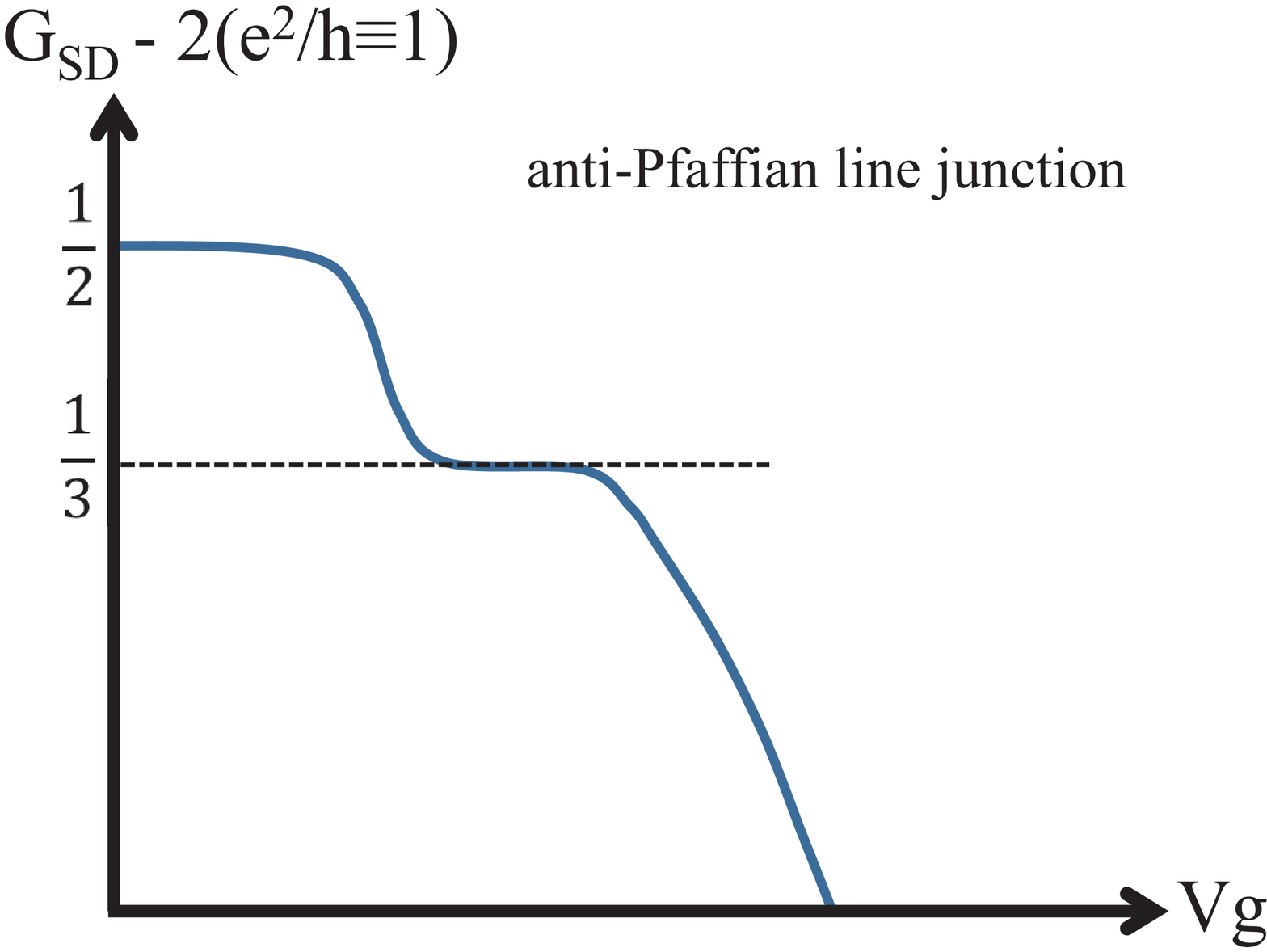}}
\caption{(Color online) Schematic plots of the conductance as a function of gate voltage $V_g$ of the long skinny gate. (a) The conductance of the Pf line junction as a function of the gate voltage $V_g$. If $V_g$ is large, the left and right edges of the line junction are well separated and the conductance is zero. Upon decreasing the gate voltage, the electron tunneling becomes stronger and the conductance starts to increase. Under RG analysis, only one stable phase exists while decreasing $V_g$ which corresponds to the phase with quantized conductance $(1/2) \frac{e^2}{h}$. (b) The conductance of the aPf line junction as a function of $V_g$. The situation is similar to (a) but upon decreasing $V_g$ there is an intermediate plateau phase with conductance, $G_{plateau}=(1/3)\frac{e^2}{h}$, smaller than the maximum quantized conductance $(1/2) \frac{e^2}{h}$.}
\label{G:5/2_line_junction}
\end{figure}
The Hamiltonian density for a clean line junction of Pf is $\mathcal{H}_0 = \mathcal{H}^{Pf}_L + \mathcal{H}^{Pf}_R + \mathcal{H}^{Pf}_{RL}$, with $\mathcal{H}^{Pf}_{R/L}$ already given in Eq.~(\ref{HPf_top_bott}) with $v^{t/b}_n\rightarrow v_{n}$, $v^{t/b}\rightarrow v_{c}$, $\psi_{t/b} \rightarrow \psi_{R/L}$ and $\phi_{t/b}\rightarrow \phi_{R/L}$. The $\mathcal{H}_{R/L}$ is the mutual interaction term
\begin{eqnarray}
\mathcal{H}^{Pf}_{RL}=4\pi \lambda n_R n_L = -\frac{\lambda}{\pi}\partial_x \phi_R \partial_x \phi_L.
\end{eqnarray}
Now the Hamiltonian of the electron tunneling between the right and left Pf edges is expressed as
\begin{eqnarray}
H_1 = \int dx \left( \eta(x) \psi_L \psi_R e^{i 2 (\phi_L - \phi_R)}+\Hc\right).
\end{eqnarray}
We assume that the random disorder is in a Gaussian distribution with $[\eta(x) \eta^*(x')]_{ens} = W' \delta(x-x')$ and introduce the non-chiral boson fields,
\begin{eqnarray}
\phi_{R/L}=\varphi \pm \frac{1}{2}\theta.
\end{eqnarray}
In terms of the new bosons, the Hamiltonian density of the clean edge line junction is $\mathcal{H}_0 = \mathcal{H}_{U(1)} + \mathcal{H}_\psi$ with
\begin{eqnarray}
&& \mathcal{H}_{U(1)} = \frac{v}{2\pi} \left[ K (\partial_x \varphi)^2 + \frac{1}{K} (\partial_x \theta)^2 \right], \\
&& \mathcal{H}_\psi = i v_n \left[ \psi_L \partial_x \psi_L + \psi_R \partial_x \psi_R \right],
\end{eqnarray}
where $v= v_c \sqrt{1-\lambda^2}$, $K= 2\sqrt{\frac{1-\lambda}{1+\lambda}}$, and the second Hamiltonian is related to the neutral Majorana fields which is not altered by introducing the new bosons fields.

The RG flow equation to the leading order $W'$ is
\begin{eqnarray}
\frac{dW'}{d\ell} = (3 - 2\Delta)W',
\end{eqnarray}
where $\Delta$ is the scaling dimension of the electron tunneling term, which can be extracted to be
\begin{eqnarray}
\nonumber \Delta&=&\Delta\left[ \psi_L \psi_R e^{i 2(\phi_L-\phi_R)}\right]=\\
&=&\Delta\left[\psi_L \psi_R e^{i 2\theta}\right] = 1+ K.
\end{eqnarray}
When $\Delta>3/2~(K>1/2)$, the electron tunneling is irrelevant and the conductance is zero at zero temperature. When $\Delta<3/2~(K<1/2)$, the electron tunneling is relevant and the left Pf and right Pf edged fuse into each other. The phase shows the quantized conductance $G-2\frac{e^2}{h}=(1/2) \frac{e^2}{h}$. The schematic plot of the conductance in the Pf line junction as a function of gate voltage $V_g$ is shown in Fig.~\ref{Pf:line_junction_G}.
\subsubsection{Anti-Pfaffian line junction, Fig.~\ref{5/2:line_junction}(b)}
\begin{figure}[t]
\includegraphics[width=\columnwidth]{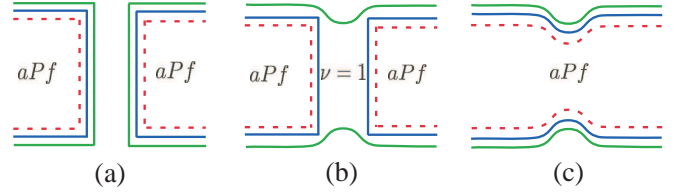}
\caption{(Color online) Schematic plots of different phases of the aPf line junction. The solid green lines are the $\nu=1$ edges, the combination of the blue lines and the dashed red lines are the Pf edge. The blue line represents a chiral boson and the dashed red line represent a neutral Majorana. (a) When gate voltage $V_g$ of the central long skinny gate is large, the left and right edges are well separated with vanishing conductance. (b) When $V_g$ decreases to a point, the electron tunneling between the left $\nu=1$ edge and right $\nu=1$ edge is relevant. Under RG, the $\nu=1$ edges fuse to become the so called plateau phase shown here. In this phase, the configuration is the same to the edge reconstruction phase of the aPf system in disorder-dominated phase, which is discussed in Sec.~\ref{Pf:Edge_recons}. The conductance in this plateau phase is expected to be $G=(1/3)\frac{e^2}{h}$. (c) When $V_g$ become too small, the Pf edges also are connected to give the quantized conductance $G-2 \frac{e^2}{h}=(1/2)\frac{e^2}{h}$.}
\label{aPfline_junction:diffstage}
\end{figure}

In the aPf line junction, the discussions are similar to those in $\nu=2/3$ line junction. For the aPf, there are more internal edge structures and an electron has more channels to tunnel across the line junction. Intuitively, there are more phase transitions in the aPf system than in the Pf system. Below, we will focus on the disorder-dominated phase in which an universal conductance are expected.

The Hamiltonian density for the aPf in the disorder-dominated phase is described by one neutral Majorana mode $\psi$ and two decoupled boson modes called charge mode, $\phi_\rho$ and neutral mode, $\phi_\sigma$. The Hamiltonian density for the aPf line junction in disorder-dominated phase in terms of boson fields and Majorana fields is $\mathcal{H}_0 = \mathcal{H}_\rho + \mathcal{H}_\sigma + \mathcal{H}_\psi$ with
\begin{eqnarray}
&&\mathcal{H}_\rho = \frac{2v_\rho}{4\pi}  \left[ (\partial_x \phi_{\rho R})^2 + (\partial_x \phi_{\rho L})^2- 2 \lambda \partial_x \phi_{\rho R} \partial_x \phi_{\rho L}\right],~~~~~~~~\\
&& \mathcal{H}_\sigma = \frac{v_\sigma}{4\pi}  \left[ (\partial_x \phi_{\sigma R})^2 + (\partial_x \phi_{\sigma L})^2\right],\\
&& \mathcal{H}_{\psi} = iv_n \left( \psi_R\partial_x\psi_R + \psi_L \partial_x \psi_L \right),
\end{eqnarray}
with $\lambda$ being the strength of Coulomb interaction between the charge modes.

Introducing the non-chiral bosons for the charge modes,
\begin{eqnarray}
\phi_{\rho R/L} = \varphi \pm \frac{1}{2} \theta,
\end{eqnarray}
we can map $\mathcal{H}_\rho$ to be Luttinger Hamiltonian density with
\begin{eqnarray}
\mathcal{H}_\rho = \frac{\tilde{v}_\rho}{2\pi} \left[ K (\partial_x \varphi)^2 + \frac{1}{K} (\partial_x \theta)^2 \right],
\end{eqnarray}
with $\tilde{v}_\rho = v_{\rho}\sqrt{1-\lambda^2}$ and $K=2\sqrt{\frac{1-\lambda}{1+\lambda}}$. With the diagonal Hamiltonian density, we can extract all the scaling dimension of interest, i.e. $\Delta[e^{i\phi_{\sigma R/L}}] =1/2$, $\Delta[e^{i \phi_{\rho R/L}}]=\Delta[e^{i (\varphi \pm \theta/2)}] = \frac{1}{4K} + \frac{K}{16}$, and $\Delta[\psi_{R/L}] =1/2$.

Now, we need to take electron tunneling between the right edge and left edge into account. An electron created on the $\nu=1$ channel of the aPf edge is represented by the operator $e^{i \phi_1}$, and an electron created on the Pf channel of the aPf edge is represented by the operator $\psi e^{-i 2 \phi_2}$.  Therefore, there are three possible electron tunneling terms we can add to the clean line junction Hamiltonian density
\begin{widetext}
\begin{eqnarray}
\nonumber \mathcal{H}_1 &= & \bigg{[} \eta_1(x) e^{i(\phi_{1R} - \phi_{1L})}+\eta_2(x) \left(\psi_R e^{-i (2\phi_{2R}+\phi_{1L})} + R\leftrightarrow L \right)+ \eta_3(x) \psi_L \psi_R e^{-i2 ( \phi_{2R}-\phi_{2L})} \bigg{]}+ \Hc \\
 & = & \bigg{[} \eta_1(x) e^{i(2\theta - \phi_{\sigma R} + \phi_{\sigma L})} + \eta_2(x) \psi_R e^{i ( 2\theta- 2 \phi_{\sigma R} + \phi_{\sigma L})} + \eta_3(x) \psi_L \psi_R e^{i 2(\theta-\phi_{\sigma R} + \phi_{\sigma L})} \bigg{]} + \Hc.
\end{eqnarray}
\end{widetext}

There are now three types of electron tunneling terms and we assume all the random disorders in the line junction are in Gaussian distribution with $[\eta_a(x)\eta_a(x')]_{ens} = W_a \delta(x-x')$, with $a=1,2,3$. As before, all of them satisfy the RG flow equations
\begin{eqnarray}
\frac{d W_a}{d \ell} = (3-2\Delta_a )W_a,
\end{eqnarray}
where $\Delta_a$ are the scaling dimensions for type-$a$ electron tunneling term. If $\Delta_a > 3/2$, the electron tunneling is irrelevant. If $\Delta_a <3/2$, the electron tunneling is relevant and the corresponding edges are fused with each other.

The scaling dimensions of each type of electron tunneling are
\begin{eqnarray}
&& \Delta \left[ e^{i ( \phi_{1L} - \phi_{1R})}\right] = 1+K,\\
&& \Delta \left[ \psi_{R/L} e^{i(2\phi_{2 R/L}-\phi_{1L/R})}\right]=3+K,\\
&& \Delta \left[ \psi_L \psi_R e^{i 2(\phi_{2R} - \phi_{2L})} \right] = 5 + K.
\end{eqnarray}
Since $0<K<2$, we can see the tunneling between the $\nu=1$ channel can be relevant once $K<1/2$. The other channels of electron tunneling are always irrelevant in this regime. Therefore, initially when the right and left edge are well separated, $K>1/2$, the electron tunneling is suppressed and conductance is zero. Upon decreasing the gate voltage to decrease the width of the line junction, the Coulomb repulsion become larger and $K$ decreases. The electron tunneling becomes stronger and the conductance starts to increase. When $K<1/2$, there is a phase transition to a phase in which the $\nu=1$ channels on the left and right edges are fused to each other. The schematic configuration of the phase evolve from Fig.~\ref{aPfline_junction:diffstage}(a) to Fig.~\ref{aPfline_junction:diffstage}(b). The phase of Fig.~\ref{aPfline_junction:diffstage}(b) is qualitatively the same to that of the edge reconstruction phase of the aPf system in Sec.~\ref{aPf:Edge_recons}. In this phase, the two-terminal conductance is expected to be $G-2\frac{e^2}{h} = (1/3)\frac{e^2}{h}$. If we keep decreasing the gate voltage, in the end the Pf channels on the left and right aPf edges are also fused to become the phase shown in Fig.~\ref{aPfline_junction:diffstage}(c). In this phase, the conductance reaches its usual quantized conductance $G-2\frac{e^2}{h} = (1/2) \frac{e^2}{h}$.

The schematic plot of the two-terminal conductance of the aPf line junction as a function of $V_g$ is shown in Fig.~\ref{aPf:line_junction_G}. The crucial feature is the existence of the plateau phase with $G-2\frac{e^2}{h} = (1/2) \frac{e^2}{h}$, which corresponds to the intermediate phase in which only parts of the edge modes are connected. Testing the existence of such a plateau phase can serve as a clear guide for distinguishing Pf from aPf state.
\bibliography{biblio4edgetpt}
 \end{document}